\documentclass[12pt,a4paper]{article}
\usepackage{jheppub}

\usepackage{graphicx}
\usepackage{subcaption}
\usepackage{xspace}
\usepackage{verbatim}

\usepackage[utf8]{inputenc}

\usepackage[style=numeric-comp,sorting=none]{biblatex}
\abstract{ A working group on detector simulation was formed as part of the high-energy physics (HEP) 
Software Foundation's initiative to prepare a Community White Paper that describes 
the main software challenges and opportunities to be faced in the HEP field over the next 
decade. The working group met over a period of several months in order to review the 
current status of the Full and Fast simulation applications of HEP experiments 
and the improvements that will need to be made in order to meet the goals of future HEP 
experimental programmes.  The scope of the topics covered includes the main components 
of a HEP simulation application, such as MC truth handling, geometry modeling, particle 
propagation in materials and fields, physics modeling of the interactions of particles 
with matter, the treatment of pileup and other backgrounds, as well as signal processing 
and digitisation. The resulting work programme described in this document focuses on the need to 
improve both the software performance and the physics of detector simulation. The goals are to increase the 
accuracy of the physics models and expand their applicability to future physics programmes, 
while achieving large factors in computing performance gains consistent with projections on 
available computing resources.}

\begin{document}

\noindent
\begin{tabular*}{\linewidth}{lc@{\extracolsep{\fill}}r@{\extracolsep{0pt}}}
 & & HSF-CWP-2017-07 \\
 & & February 28, 2018 \\ 
 & & \\
\end{tabular*}
\vspace{0.1cm}

\title{HEP Software Foundation Community White Paper Working Group -- Detector Simulation}

\author[c]{J Apostolakis}
\author[p]{M Asai}
\author[e]{S Banerjee}
\author[l]{R Bianchi}
\author[e]{P Canal}
\author[k]{R Cenci}
\author[b]{J Chapman}
\author[c]{G Corti}
\author[c]{G Cosmo}
\author[o]{S Easo}
\author[h]{L de Oliveira}
\author[p]{A Dotti}
\author[e,1]{V Elvira}
\author[h]{S Farrell}
\author[e]{L Fields}
\author[e]{K Genser}
\author[c]{A Gheata}
\author[c]{M Gheata}
\author[c,1]{J Harvey}
\author[c]{F Hariri}
\author[e]{R Hatcher}
\author[e]{K Herner}
\author[j]{M Hildreth}
\author[c]{V Ivantchenko}
\author[e]{T Junk}
\author[e]{S-Y Jun}
\author[e]{M Kirby}
\author[m]{D Konstantinov}
\author[e]{R Kutschke}
\author[e]{P Lebrun}
\author[e]{G Lima}
\author[e]{A Lyon}
\author[h]{Z Marshall}
\author[c]{P Mato}
\author[a]{M Mooney}
\author[p]{R Mount}
\author[i]{J Mousseau}
\author[h]{B Nachman}
\author[e]{A Norman}
\author[c]{M Novak}
\author[e]{I Osborne}
\author[q]{M Paganini}
\author[e]{K Pedro}
\author[c]{W Pokorski}
\author[a]{X Qian}
\author[e]{J Raaf}
\author[k]{M Rama}
\author[c]{A Ribon}
\author[c]{S Roiser}
\author[n]{D Ruterbories}
\author[e]{S Sekmen}
\author[g]{B Siddi}
\author[e]{E Snider}
\author[f]{S Vallecorsa}
\author[d]{M Verderi}
\author[e]{H Wenzel}
\author[c]{S Wenzel}
\author[a]{B Viren}
\author[e]{T Yang}
\author[e]{J Yarba}
\author[e]{L Welty-Rieger}
\author[p]{D H Wright}
\author[a]{C Zhang}

\vspace{0.5cm}

\affiliation[a]{Brookhaven National Laboratory, Upton NY, USA}
\affiliation[b]{University of Cambridge,Cambridge, UK}
\affiliation[c]{CERN, Geneva, Switzerland}
\affiliation[d]{IN2P3-LLR, Ecole Polytechnique, Palaiseau, France}
\affiliation[e]{Fermilab National Accelerator Laboratory, Batavia IL , USA}
\affiliation[f]{Gangneung-Wonju National University, Gangneung, Gangwon-do, South Korea}
\affiliation[g]{Università di Ferrara and INFN Ferrara, Ferrara, Italy}
\affiliation[h]{Lawrence Berkeley National Laboratory, Berkeley CA, USA}
\affiliation[i]{University of Michigan, Ann Arbor MI, USA}
\affiliation[j]{University of Notre Dame, Notre Dame, IN 46556, USA}
\affiliation[k]{Università di Pisa and INFN Pisa, Pisa, Italy}
\affiliation[l]{University of Pittsburgh, Pittsburgh PA, USA}
\affiliation[m]{Institute For High Energy Physics(IHEP), Protvino,Russia}
\affiliation[n]{University of Rochester, Rochester NY, USA}
\affiliation[o]{Rutherford Appleton Laboratory, Chilton, UK}
\affiliation[p]{SLAC National Accelerator Laboratory, Menlo Park, CA 94025, USA}
\affiliation[q]{Yale University, New Haven, CT 06520, USA}
\affiliation[1]{Paper Editor}

\maketitle

\newpage

\hypertarget{introduction}{%
\section{Introduction}\label{introduction}}

Detector simulation is an essential tool to design, build and commission
the highly sophisticated detectors utilised in modern particle physics
experiments and cosmology. It is also a fundamental tool for data
analysis and interpretation of the physics measurements. The typical
simulation software application in high-energy physics (HEP) experiments
consists of a set of packages executed in sequence, usually starting
with a generator of the physics processes under study. In particle
colliders, event generators provide the final state particles observed
as primary particles in the detector. A second module simulates the
passage of the generated primary particles through the detector material
and its magnetic field. In most contemporary experiments, this detector
simulation module is based on the Geant4 simulation toolkit, a software
package that provides the tools to describe the detector geometry and
materials, and incorporates a number of models to simulate
electromagnetic and hadronic interactions of particles with matter~\cite{ALLISON2016186}. 
When appropriate, various fast simulation options can be used
to save computing time with some penalty on physics accuracy. The next
module, often called digitisation, simulates the readout process,
including the signal processing by the detector electronics, as well as
the calibration of individual channels. At the end of the chain, the
same algorithms used to identify and reconstruct individual particles
and physics observables in real data are applied to simulated events.

In the case of accelerator-based neutrino experiments, the chain
typically starts with a Geant4 module to model the interaction of a
proton beam with a target to produce a neutrino beam which is driven to
a detector where the physics interaction of interest occurs. The
neutrino beam and the first neutrino-nucleus interaction in the detector
is typically modeled with the GENIE ~\cite{Andreopoulos:2009rq} package, while Geant4 is
utilised in a second step to simulate the nucleus de-excitation and
subsequent hadron production. Readout modeling, calibration, and
reconstruction is done as described for collider experiments.

During the last three decades, simulation has proven to be of critical
importance to the success of HEP experimental programmes. For example,
the detailed detector modeling and accurate physics of the CERN Large
Hadron Collider (LHC) CMS and ATLAS experiments Geant4-based simulation
software helped these collaborations deliver physics results of
outstanding quality faster than any previous hadron collider experiment.
The FLUKA simulation package has also been used extensively for studies
of shielding and radiation modeling, providing important input to the
design of the original LHC experiments and their upgrades. Simulation
software at the LHC experiments is more accurate and yet runs much
faster than its predecessor at the Tevatron. Simulation samples of
better quality and in larger quantities, evolving detector and computing
technology, and a wealth of experience from pre-LHC experiments on
calibration and data analysis techniques, improved significantly the
precision of the measurements in the current generation of experiments.
In parallel with this, simulation has been used to design a new
generation of neutrino and muon experiments and in searches for beyond
the standard model (BSM) physics in precision experiments.

For all its success so far, the challenges faced by the HEP field in the
simulation domain are daunting. During the first two runs, the LHC
experiments produced, reconstructed, stored, transferred, and analyzed
tens of billions of simulated events. This effort required more than
half of the total computing resources allocated to the experiments. As
part of the high-luminosity LHC physics program (HL-LHC) and through the
end of the 2030's, the upgraded experiments expect to collect 150 times
more data than in Run 1. Demand for larger simulation samples to satisfy
analysis needs will grow accordingly. In addition, simulation tools have
to serve diverse communities, including accelerator-based particle
physics research utilizing proton-proton colliders, neutrino and muon
experiments, as well as the cosmic frontier. The complex detectors of
the future, with different module- or cell-level shapes, finer
segmentation, and novel materials and detection techniques, require
additional features in geometry tools and bring new demands on physics
coverage and accuracy within the constraints of the available computing
budget. More extensive use of Fast Simulation is a potential solution,
under the assumption that it is possible to improve time performance
without an unacceptable loss of physics accuracy.

The effort to improve detector simulation and modeling tools including
the accuracy of their physics, whilst saving memory allocation and
computing execution time, requires immediate attention. This report
addresses the future development of Full Simulation and Fast Simulation
applications of HEP experiments for detectors and beam lines with a view
to achieve improvements in software efficiency, scalability and
performance, and to make use of the advances in processor, storage and
network technologies. A plan must consider new approaches to computing
and software that could radically extend the physics reach of the
detectors, while ensuring the long-term sustainability of the software
through the lifetime of the experiments running through the 2020's and
2030's. One example is the GeantV ~\cite{1742-6596-396-2-022014} R\&D project, started in 2013.
Another example is the ongoing effort to port simulation code to run on
High Performance Computing (HPC) facilities, which provide massive CPU
power and modern computing architectures.

Components of the HEP simulation chain, such as physics generators and
event visualisation tools are discussed elsewhere, while the
modeling of the propagation and beam manipulation through accelerator
lattices (i.e. CERN and Fermilab accelerator complexes) are considered
to be outside the scope of this report. Space and medical applications
are referred to briefly where appropriate.


\hypertarget{challenges}{%
\section{Challenges}\label{challenges}}

The experimental programmes planned in the next decade are driving
developments in the simulation domain; they include the High Luminosity
LHC project (HL-LHC), neutrino and muon experiments, and studies towards
future colliders such as the Future Circular Collider (FCC) and the
Compact Linear Collider (CLIC). The requirement of improving precision
in the simulation implies production of larger Monte Carlo samples,
which scale with the number of real events recorded in future
experiments, and this places an additional burden on the computing
resources that will be needed to generate them. The diversification of
the physics programmes also requires new and improved physics models.

The main challenges to be addressed if the required physics and software
performance goals are to be achieved can be summarised as follows:

\begin{itemize}
\item
  reviewing the physics models' assumptions, approximations and
  limitations in order to achieve higher precision, and to extend the
  validity of models up to energies of the order of 100 TeV foreseen
  with the Future Circular Collider (FCC) project;
\item
  redesigning, developing, commissioning detector simulation toolkits to
  be more efficient when executed on current vector CPUs and emerging
  new architectures, including GPGPUs where use of SIMD vectorisation is
  vital; this includes porting and optimising the experiments'
  simulation applications to allow exploitation of large HPC facilities;
\item
  exploring different Fast Simulation options, where the full detector
  simulation is replaced, in whole or in part, by computationally
  efficient techniques; an area of investigation is common frameworks
  for fast tuning and validation;
\item
  developing, improving, and optimising geometry tools that can be
  shared among experiments to make the modeling of complex detectors
  computationally more efficient, modular, and transparent;
\item
  developing techniques for background modeling, including contributions
  of multiple hard interactions overlapping the event of interest in
  collider experiments (pile-up);
\item
  revisiting digitisation algorithms to improve performance and
  exploring opportunities for code sharing among experiments;
\item
  recruiting, training, retaining, human resources in all areas of
  expertise pertaining to the simulation domain, including software and
  physics.
\end{itemize}

The complexity and diversity of future experiments results in an
enormous demand for CPU, and detailed detector simulation already
consumes a large fraction of CPU resources of most current HEP experiments.
Some progress has already been made in improving software performance,
for example by adding multithreading capability to Geant4, and by
redesigning the geometry library into the VecGeom package that exploits
vectorisation. These R\&D programmes need to continue in order to
benefit from the various forms of parallelism offered by modern
architectures. For example, the goal of the GeantV R\&D project is to
build a realistic prototype application for a complex experiment with
the aim of demonstrating a big improvement in CPU time. The main
architectural feature of this prototype is the implementation of
particle-level, i.e. track-level, parallelisation for more efficient use
of SIMD vectorisation and data locality. Whenever possible, new
developments resulting from the R\&D programme will be tested in
realistic prototypes and then be integrated, validated, and deployed in
a timely fashion in Geant4 in order to maximise the benefits to
experiments at the earliest opportunity.

A flexible simulation strategy is needed that is capable of delivering a
broad spectrum of options offering different levels of physics accuracy,
from a pure Full Simulation approach to a combination of Geant4 showers,
parametrised response functions, and shower libraries. R\&D in the realm
of Fast Simulation includes the use of Machine Learning (ML) techniques
and embedded tools for tuning and validation in an effort to provide
faster and more accurate physics and to be able to safely expand the use
of Fast Simulation in the context of limited computing resources.

HPC facilities offer a valuable resource and work must continue to adapt
simulation applications to run efficiently on these systems, which may
produce a significant fraction of the simulation samples in future HEP
experiments.

Work also needs to continue to improve the accuracy of the physics
models, and the validity of the models needs to be verified for new
detectors that include, for example, liquid argon time projection
chambers (for DUNE) and silicon-absorber calorimeters (for CMS). The
models need to be improved and validated for a large variety of incident
particles and energy ranges, including muons, neutrinos, and the high
energies accessible in a Future Circular Collider (FCC).

In producing new models of physics interactions and improving the
fidelity of the models that exist, it is absolutely imperative that
high-quality data is available. Simulation model tuning often relies on
test beam data, and a program to improve the library of available data
could be invaluable to the community. Such data would ideally include
both thin-target test beams for improving interaction models and
calorimeter targets for improving shower models. These data could
potentially be used for directly tuning fast simulation models, as well.
At the least it is important that the experiments carefully consider
what measurements could be made that would be useful input to the
community of Monte Carlo program developers and tuners.

High pile-up simulation for the high-luminosity environment in future
data-taking at the LHC poses challenges in memory utilisation and data
transfer (I/O) for generating the large statistics samples needed to
compare with data from high trigger rates. It currently necessitates the
generation and handling of large minimum bias samples in order to
achieve an accurate description of high pile-up collision events. It
would be desirable to have tools that allow overlay of both simulated
and real data events. The latter poses a real challenge related to the
collection of unsuppressed zero-bias data samples that reflect the
luminosity profile of the physics run.

The digitisation component of the simulation chain offers opportunities
for parallelisation, given that the code between sub-detector components
is typically orthogonal. In the case of neutrino detectors based on
liquid argon technology, this step is computationally expensive, as
compared to other elements of the chain. Since digitisation is often
either memory- or I/O-intensive, there is a challenge to ensure that
parallelism is efficient. Another issue is that digitisation code,
including containers, data types and data structures, is typically
developed by the user. In order to exploit the SIMD opportunities
intrinsic to the digitisation process, digits must be constrained to a
subset of the possible data structures, with priority given to data
contiguity.

New areas of commonality across experiments need to be explored, for
example in the domains of Fast Simulation and geometry. A great deal of
common code is relied upon by many users for the benefit of the whole
HEP community for the long term. It will be important to identify these
pieces of code, remove the experiment dependency where appropriate, and
ensure that the pieces that cannot be removed are either going to be
maintained by a group that feels responsibility for them or will not
become a bottleneck for any detector simulation in the future.

The large reliance on simulations implies the need for careful
assessment of the systematic uncertainties associated to Geant4
predictions. New capabilities and tools are being developed to allow for
the calculation of systematic uncertainties on the physics observables
predicted by the Geant4 beam and detector simulations. In other words,
kinematic distributions, efficiencies, and calibration constants will be
predicted with systematic uncertainties resulting from the propagation
of the uncertainties associated with the individual Geant4 physics
models.

There are specific challenges associated with the Intensity Frontier
experimental programme. Properly simulating a neutrino experiment like
NOvA ~\cite{Adamson:2016tbq}, MicroBooNE~\cite{MicroBooNE}, 
MINERvA~\cite{Aliaga:2013uqz}, or DUNE ~\cite{Acciarri:2015uup}
involves a three-part software stack, the first and last of which are
relevant to this paper; the second, event generators such as GENIE, 
is discussed elsewhere. The first element of the stack concerns
the beamline and the neutrino flux prediction. Estimating the neutrino
flux is a notorious problem because as weakly interacting particles,
neutrinos offer no independent mechanisms for measuring the flux. A
central problem is in the difficulty of accurately simulating the
hadronic physics of meson production, especially in a thick target.
Additionally, the simulated neutrino flux depends very strongly on
minute details of the beamline, meaning that neutrino beam kinematic
distributions must be described in great detail. Geant4 is the most
commonly used software toolkit for this work, although FLUKA ~\cite{Ferrari:2005zk} 
and MARS ~\cite{MARS} are also used. The final element of the stack is the
detector simulation. Again there are numerous challenges in terms of
efficiently simulating large numbers of events inside complex detector
geometries and in handling subtle physics effects in the traversal of
radiation across matter. Neutrino experiments in particular rely heavily
on detector simulations to reconstruct neutrino energy, which requires
accurate modeling of energy deposition by a variety of particles across
a range of energies. Geant4 is the universal solution at this stage.

Muon experiments such as Muon g-2 and Mu2e also face large simulation
challenges, as Geant4 is the primary simulation toolkit for
all stages of these experiments. Because they are searching for extremely rare effects, they
must grapple with very low signal to background ratios, and the modeling
of low cross-section background processes. Often the physics of these
processes is not well understood, introducing large systematic
uncertainties. Additionally, the size of the computational problem is a
serious challenge, as large simulation runs are required to adequately
sample all relevant areas of experimental phase space, even when
techniques to minimise the required computations are used. Another
aspect related to the background events is the need to simulate the
effects of low energy neutrons, which requires large computational
resources. For example, Mu2e used a total of approximately 60 million
CPU hours, or 83 thousand core months, in a 2016 simulation campaign to
produce the Full Simulation sample needed to complete a technical design
report delivered in 2016. 

Although in comparison the current computing needs of neutrino and muon experiments
are small compared to the LHC experiments, i.e. the CMS experiment
spent 860 thousand core months on simulation in the May 2015-May 2016
period, they are expected to grow substantially in the next decade.
It is therefore imperative that the
intensity frontier experiments make use of modern computing
architectures and algorithms. This transition will be complicated by the
fact that the intensity frontier is fractured into many small
experiments, some of which have as few as 50 collaborators, limiting the
manpower available for updating detector simulations to use new tools.
Efforts such as LArSoft~\cite{LArSoft} have recognised this challenge and 
aim to pool resources of many experiments for the benefit of all.

\hypertarget{geometry}{%
\section{Geometry}\label{geometry}}

The geometry modeler is a key component in Monte Carlo detector
simulation, combining both solids modelling and efficient navigation
techniques. Stringent requirements apply to provide the right level of
flexibility, robustness and accuracy in modeling from the simplest to
the most complex setups and to offer adequate precision and efficiency
in the implemented algorithms. Geant4 provides a wide variety of
tools and solutions for describing geometry setups from simple to highly
complex. Geometrical models of the LHC detectors, for instance, easily
reach millions of geometrical elements of different kinds combined
together in hierarchical structures. The geometry modeler provides
techniques by which memory consumption can be greatly reduced, allowing
regular or irregular patterns to be easily replicated and assembled.
This, combined with navigation and optimisation algorithms, allows the
efficient geometry computation of the simulated tracks with the elements
composing any geometry setup.

Recent extensions of the geometry modeler in Geant4 include specialised
navigation techniques and optimisation algorithms to aid also in medical
simulation studies. This has allowed complex geometrical models of the
human body to be developed. Extensions also include parallel navigation
and tracking in layered geometries which allow geometry setups to be
superimposed on one another with minimal impact on CPU time (Figure 1).
An important and rather useful construct for shapes delimited by any
kind of complex surface is offered by the tessellated-solid construct,
which allows complex geometrical shapes to be described by approximating
their surfaces as a set of planar facets (triangles), with tuneable
resolution. This technique is used nowadays for importing geometries
from CAD systems to generate surface-bounded solids. Recent developments
introduced in Geant4 along with the implementation of the Unified Solids
library ~\cite{1742-6596-396-5-052035} provide considerably improved CPU performance, making
it possible to use such constructs for very detailed and realistic
descriptions of surfaces, while optimally scaling with the number of
facets in use. Further improvements to the tessellated solid construct
are expected as part of the rewrite of most geometry primitives going on
in the VecGeom ~\cite{1742-6596-608-1-012023} package, initiated within 
the GeantV project.

The Unified Solids Library started as an AIDA project in 2011, with the
aim to unify and improve the algorithms provided by the Geant4 and ROOT
~\cite{Brun1996} approaches to solids modeling. The library which comes with a
large collection of primitives, has been extended and enhanced as part
of the AIDA-2020 project, and is now integrated in VecGeom, which
provides an independent geometry modeler library for use as alternative
replacement in either Geant4 or ROOT. In VecGeom, the interfaces to
geometrical primitives have been extended to support vector signatures,
which are essential for GeantV, and allow for parallel processing of
queries; algorithms are completely reviewed and in good part rewritten
to make use of SIMD instructions wherever possible, and support
different kernels and architectures, including accelerators such as
Intel Xeon Phi® and GPU devices. The code has been reengineered to make
use of C++ template techniques, allowing for more modular implementation
and for specializing efficiently on different topologies.

The benefits in CPU time are clearly visible with ray-tracing scalar
tests and also measurable in terms of few percent when full physics and
electromagnetic field integration is applied.

The VecGeom geometry modeler includes a highly efficient navigation
system. It has been developed and optimised to exploit vectorised
transport on CPU, GPGPUs and Intel Xeon Phi, making use of the full
potential of the vectorised features of the library. It is planned in
the near future to provide hooks within both Geant4 and ROOT for using
the navigation system of VecGeom, eventually boosting speed further also
for the scalar mode.

\begin{figure}[bthp]
\vspace*{0.3cm}
\centering
\includegraphics[width=0.94\textwidth]{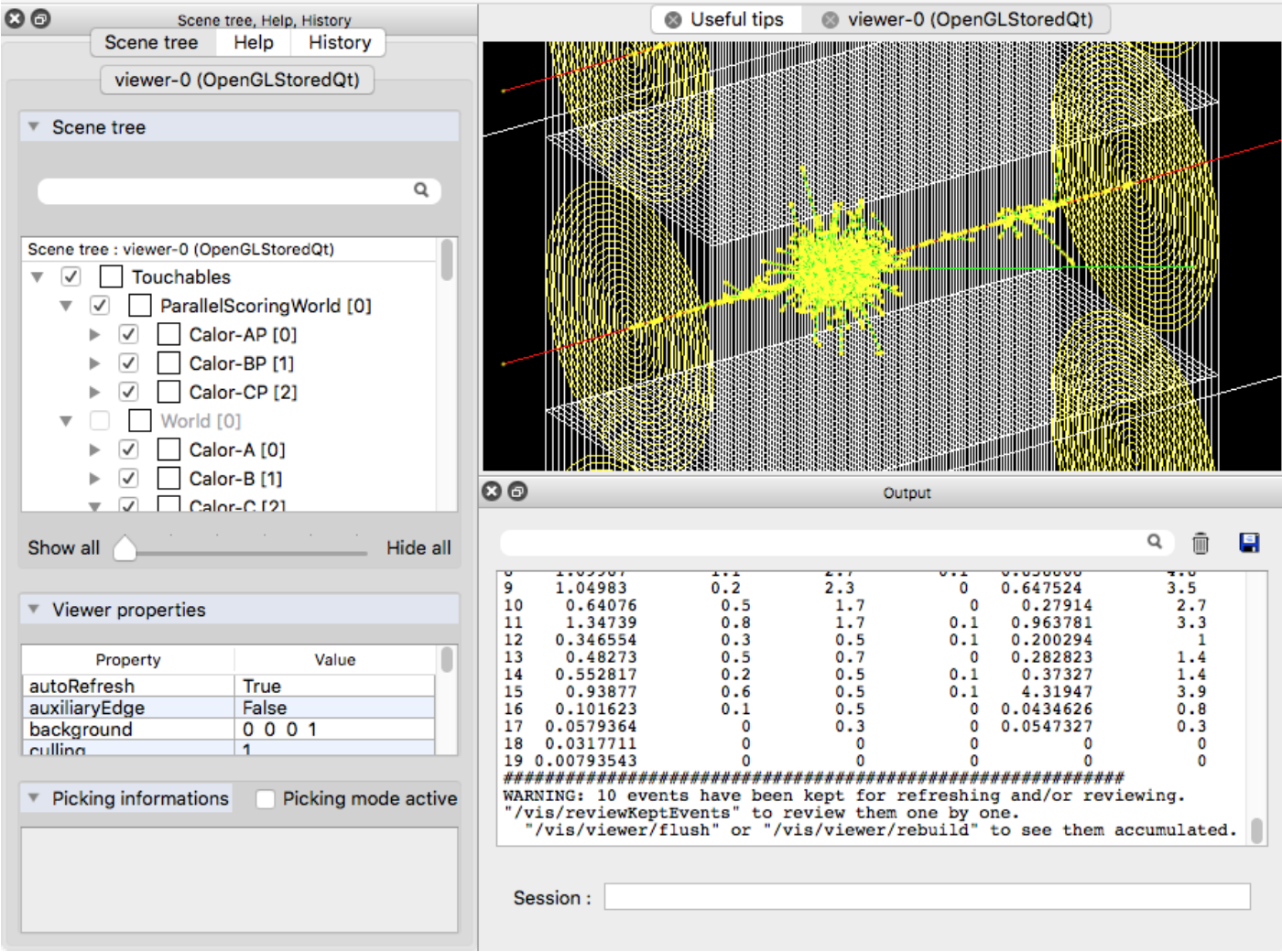}
\caption{An interactive Geant4 example featuring parallel
geometries and scoring with VecGeom primitives and parameterisations}
\label{fig:geom}
\end{figure}

Geant4 offers the ability to import and export Detector geometrical
descriptions from text files according to the Geometry Description
Markup Language (GDML) ~\cite{1710291} based on XML; the exchange format is
also supported by ROOT and the schema includes all the geometrical
primitives currently in use. GDML supports the modularisation of
geometry descriptions to multiple GDML files, allowing for rational
organisation of the modules for complex setups. Verification of the GDML
file against the latest version of the schema comes for free thanks to
Xerces-C++, with the possibility to turn it on or off in the
Geant4 GDML parser.

The DD4hep Geometry Toolkit ~\cite{1742-6596-513-2-022010}, being developed in the context of
the AIDA and AIDA-2020 projects aims to provide a generic detector
description toolkit built by reusing the existing components of widely
used packages and provide the missing functional elements and
interfaces, for offering a complete and coherent detector description
solution to experiments.

Computations performed by a geometry modeler in a simulation program
must be highly precise and efficient. Lots of effort has been spent in
the past decades to achieve the level of precision and reliability
offered today, still the evolution of the computing hardware, compilers
and programming languages require algorithms and related implementations
to evolve as well. It is therefore foreseeable that effort in this sense
will have to continue, in a way as it is currently being done for the
VecGeom package, to also favor modularity and reusability of the various
components.

A better integration of the existing tools used in simulation and the
adoption of CAD systems for designing and fast-prototyping the geometry
models is also desirable. A variety of solutions tailored for simulation
exist today, especially in tools developed for space science
applications, although none or few of them are fully open source. In the
past such a system was not desirable because of the difference in level
of detail required in CAD programs compared to standard detector
simulations. Individual pipes, bolts, and wires, often described in CAD,
are normally grouped together and modeled as single large volumes in
detector simulations. In some modern detectors, the precision of the
physics program necessitates a level of detail in the detector geometry
that is more comparable to that of a CAD design model.

Several of the systems for geometry description, while useful for
detector simulation, lack important features for use in data
reconstruction. The most important of these features is the ability to
include detector (mis-)alignment effects, some of which may cause
collisions between the 3-dimensional volumes in the detector. It is
critical to include these effects in reconstruction software, but often
in detector simulation such small alignment issues are omitted.

The programme of work for the near future in the field of geometry
includes the following:

\begin{itemize}
\item
  Promote the adoption of VecGeom by the experiments and users
  community, by adiabatically integrating and validating its use through
  the interfaces now existing and provided in the latest versions of the
  Geant4 toolkit.
\item
  Complete the porting and implementation in VecGeom of the missing
  shapes from the standard set of primitives of the GDML schema;
  deprecate the old scalar USolids implementation and tidy up code and
  configuration settings in VecGeom.
\item
  Provide VecGeom documentation and user guides.
\item
  Investigate adoption of VecGeom navigation algorithms in Geant4 and
  ROOT and provide first prototype implementations of a navigator
  adapter making full use of the scalar navigation capabilities in
  VecGeom.
\item
  Provide implementation of a GDML reader/writer in VecGeom to allow for
  easy detector description geometry exchange.
\end{itemize}

\hypertarget{digitisation}{%
\section{Digitisation}\label{digitisation}}

Simulation toolkits like Geant4, in the end, only provide energy
depositions positioned in space and time in geometric elements. These
simulators often do not include some effects like charge drift in an
electric field, which would be prohibitively expensive, and often
unnecessary, to model in detail, and they do not include models of the
readout electronics of the experiments. They may not even have knowledge
of the detailed readout structure of the detector, for example
simulating charged particles passing through a block of silicon rather
than individual pixels or strips that correspond to readout channels.
Instead, these effects are normally taken into account in a separate
step called digitisation. The input to digitisation is the ``MC truth
information'' and energy deposits from the simulation, and the output
often includes a new, processed truth record, as well as a data format
that is conceptually similar to the data from the detector (e.g.
voltages, currents, times, and so on).

Digitisation is inherently local to a given sub-detector, and often even
to a given readout element, so that there are many opportunities for
parallelism if the code and the data objects are designed optimally.
Digitisation is also normally the stage at which pileup is introduced
and backgrounds, including detector noise, are included. Generally,
without pileup, the digitisation process is not particularly expensive
in terms of CPU time or memory. With pileup included, however, there can
be significant performance issues, particularly for HL-LHC
configurations of digitisation code. One exception to this performance
rule are large drift detectors like the ALICE TPC, where significant
``simulation'' occurs in the digitisation software. Again, parallelism
can provide a path to performance improvements, both in terms of
vectorisation and multiprocessing or multithreading.

Even without major performance challenges, digitisation presents one of
the most significant risks in terms of code aging and loss of expertise.
The detailed understanding of the readout structure of a detector is
normally most readily available when the detector is built, and as time
goes on the experts in these readout systems may move on, making code
maintenance and development, particularly for detector upgrades,
problematic. Similarly, because these codes often do not result in major
and obvious analysis-level issues, they are often left alone for many
years, without being updated to newer standards and best practices. Such
issues could be helped by greater code sharing, if it could be achieved.

Since digitisation code includes a number of effects that are specific
to the individual readout that is used, historically there has been very
limited sharing of code among digitisation frameworks. Recently, both
hardware and software projects have benefitted from an increased level
of sharing among experiments. LArSoft is an endeavor to provide common
digitisation code to all the experiments using large liquid argon
detectors. Similarly, the development of next generation silicon
detectors requires realistic simulation of the charge collection and
digitisation processes. Owing to the large variety of technologies,
common software frameworks need to be flexible and modular to cater the
different needs. The Allpix software and the more recent Allpix Squared
framework provide generic algorithms shared between many silicon pixel
detector projects, and Allpix Squared follows a modular approach and is
thus also suited to simulate silicon-based calorimeters ~\cite{Allpix}. The
frameworks also allow the implementation of effective radiation damage
models and therefore enable the simulation of the detector response
throughout the experiment lifetime.

As both CMS and ATLAS expect to use similar readout chips in their
future trackers, further code sharing might be possible. Such effects
will be more important in the coming years, and the experiments both at
the LHC and elsewhere are examining similar detector technologies for
future upgrades or implementations, so the possibility of common code
bases should not be ignored.

In summary, it is expected that within the next 3 years common
digitisation efforts are well-established among experiments, and
advanced high-performance generic digitisation examples, which
experiments could use as a basis to develop their own code, will become
available. For example, the development of next generation silicon
detectors requires realistic simulation of the charge collection and
digitisation processes. Owing to the large variety of technologies,
common software frameworks need to be flexible and modular to cater for
the different needs. Specific objectives are by:

\begin{itemize}
\item
  2020, to deliver advanced high-performance, SIMD-friendly generic
  digitisation examples that experiments can use as a basis to develop
  their own code.
\item
  2022, to fully test and validate optimised digitisation code that can
  be used by the HL-LHC and DUNE experiments.
\end{itemize}

\hypertarget{modeling-of-pileup}{%
\section{Modeling of Pileup}\label{modeling-of-pileup}}

For the LHC experiments, in particular ATLAS and CMS, the modeling of
pileup (proton-proton collisions simultaneous with the collision of
interest) represents a significant challenge over the next 10 years. The
amount of pileup is expected to grow from an average of
\textasciitilde{}30 collisions for every proton-proton bunch crossing in
2016, to \textasciitilde{}200 in the HL-LHC era in 2025. The individual
detector elements of the experiments are sensitive to pileup in
different ways, and over different time ranges. It is perhaps worth
noting that the discussion of pileup here can equally apply to the
high-multiplicity underlying event in central heavy-ion collisions at
the LHC.

There are a number of ways to model the effect of pileup that the
experiments have all examined. The simplest approach is to simulate a
large number of inclusive proton proton collision (minimum bias) events
with an event generator, such as Pythia8 ~\cite{Sjostrand:2014zea}, and to overlay the
energy depositions from these events on top of the energy depositions
from the event of interest. Normally these minimum bias events are kept
in libraries to avoid frequent, CPU intensive re-simulation of all the
required pileup. With the pileup expected at the HL-LHC, these libraries
become very large if the reuse rate of events with high-pT jets is to be
kept small, demanding large amounts of disk storage for each job as they
copy a sufficient number of events locally. The copying introduced also
strains networks when large numbers of jobs are running.

Another option is to ``pre-mix'' together the minimum bias collisions
into individual events that have the full background expected for a
single collision of interest. This has the advantage that the disk usage
of a single event is reduced, but still runs a risk of background events
being reused for a large dataset.

The experiments have also explored the possibility of using data
directly to model the pileup and detector noise, a technique called
``overlay". This option has a number of advantages, including the
correct modeling of detector noise over the full run period. There are,
however, a few drawbacks that make this both technically and practically
difficult. One can only model detectors for which data has been taken,
so simulations of future layouts cannot be done this way. The detector
misalignments and simulation misalignments may not be the same, but the
events are normally run through a single reconstruction, so that hits
shift positions if different alignments are used for both. The data used
for overlay must also normally be taken without ``zero suppression'' in
order to ensure that low energy contributions from both the signal event
and background that are individually below the normal readout threshold
for a detector can coherently add to a deposition above that threshold.
In addition, some detector electronics have nonlinear responses to
deposited energy, implying that their output data cannot be added
trivially. Complications and approximations could be necessary in a data
overlay scheme to deal with nonlinear electronics.

With very fast simulation come additional possibilities. For
sufficiently fast simulation, it becomes feasible to perform the
simulation (and even the generation) of the additional pileup events at
the same time as the signal event. This ensures that the overlaid events
are unique, but has the disadvantage of the fast simulation's lower
fidelity. Under normal circumstances, however, analyses are not
sensitive to the details of the pileup simulation, and even a fast
simulation that is not sufficiently accurate for a signal event
simulation may be sufficient when simulating the pileup. Indeed, with
generative adversarial networks such as those discussed in the Fast
Simulation Section 10.3, the background from pileup could be generated
as a single background image, taking into account all backgrounds that
don't require special handling. Such techniques are obvious candidates
for exploration in the coming years.

Both experiments and theorists have also explored the possibility of
parameterisations of either the pileup or the detector response to the
pileup. Programs like DELPHES and PGS have simple parameterisations of
the response and resolution of analysis objects, and in general the
impact of pileup can be included, at least approximately, by modifying
those resolutions and lowering identification and isolation efficiencies
for leptons. Such techniques are rough, but are often good enough for
analysis prototyping or for far-future feasibility studies before the
experiments invest significant effort in the detailed simulation that
might be needed for the analysis. Such techniques are often employed
when performing large signal parameter space scans, as in the case of
searches for Supersymmetry, before detailed simulation of specific
signal models is undertaken.

In the parametric approach, any fast simulation technique used to model
the pileup must have a sufficiently correct model of the physics of
interest for the analysis at hand. For some analyses of long-lived
particles, this may not be the case, so multiple solutions should always
be kept in play.

As the LHC experiments further develop their data overlay techniques, it
may be useful to hold occasional discussions of challenges overcome in
these systems. Alignment issues, data-collection issues, and so on can
be addressed with the same philosophical approach, even if the technical
implementations differ somewhat. The development of fast simulations for
pileup should be followed as a part of the fast simulation discussions
explored in Section 10.3. As studies of physics analyses in
high-luminosity environments become more common, some of the current
parameterisation and fast simulation techniques will improve their
fidelity, but this will happen as a natural evolution with
experimentalists and phenomenologists, and likely does not require any
directed action.

In summary, an R\&D programme is needed to study different approaches to
managing the various backgrounds to hard-scattering events, including
in-time pileup, out-of-time pileup, cavern background and beam-gas
collisions. Progress is expected on the following within the next 3
years:

\begin{itemize}
\item
  Real zero-bias events can be collected, bypassing any zero
  suppression, and overlaid on the fully simulated hard scatters. This
  approach faces challenges related to the collection of
  non-zero-suppressed samples or the use of suppressed events,
  non-linear effects when adding electronic signals from different
  samples, and sub-detector misalignment consistency between the
  simulation and the real experiment. Collecting calibration and
  alignment data at the start of a new Run would necessarily incur
  delays such that this approach is mainly of use in the final analyses.
  The experiments are expected to invest in the development of the
  zero-bias overlay approach by 2020.
\item
  The baseline option is to ``pre-mix'' together the minimum bias
  collisions into individual events that have the full background
  expected for a single collision of interest. Experiments will invest
  effort on improving their pre-mixing techniques, which allow the
  mixing to be performed at the digitisation level reducing the disk and
  network usage for a single event.
\end{itemize}

\hypertarget{handling-of-monte-carlo-truth-information}{%
\section{Handling of Monte Carlo Truth
Information}\label{handling-of-monte-carlo-truth-information}}

All simulations keep some concept of Monte Carlo Truth, a record of what
was really going in the simulation of the detector, unaltered by the
detector readout. This is in some cases as simple as the event record
from the generator, which includes the particles that were input to the
simulation as well as some of the upstream particles from which they
derive. In most cases, however, more complex MC Truth is desired.
Interestingly, the standards for these records exist as text formats and
as C++ classes (HepMC and LHE), but there are not widely-used compressed
binary storage formats for the classes. Having such a standard might
facilitate the reusability of code and help the examples provided with
simulation toolkits be more practically useful.

One standard addition to the event generator record is a record of a
limited set of particle interactions in the detector. Often, for
example, the hadronic interaction of a pion within the tracker volume is
stored. There is not a standard format for such interactions, but it is
common to add these interactions directly to the original event
generator record, which is normally stored as a connected tree. This
tree is unique for a given event, but if multiple particles are
simultaneously in flight, as in the case of highly-parallel simulation,
then they may need to make simultaneous modifications to this tree. This
problem is not insurmountable, but requires some care.

The various other types of MC Truth are often detector-specific. These
can include records of true energy deposition, as well as records of
which particles in the generator record resulted in those energy
depositions, for example. Such records are more straightforward to store
in tracking detectors, where normally there is a one-to-one
correspondence between a ``hit'' used for track reconstruction and an
energy deposition by a particle from the event generator. Calorimeter
records are inherently more complicated, as showers from many particles
can overlap and merge.

The most significant challenge for MC Truth handling is keeping the
memory and disk consumption minimal, while still storing all the
information that is important to the analysis groups downstream. For
example, Geant4 has a concept of a ``Trajectory'', which can be useful
for keeping a particle history as a part of the truth record. If such
trajectories are kept for all particles produced in the calorimeter of
an LHC experiment, however, then the increase in memory usage would
become prohibitive. Similarly, while most interactions can be stored in
the truth record in the tracker, it would not be practical to do so in
the calorimeter, where each event results in millions of secondaries
generated. Ensuring that each experiment has the flexibility to save the
required MC Truth information, without inducing significant overheads in
memory or CPU usage, is critical for any future simulation toolkit.

For backgrounds that are included, as in the case of pileup, it is also
important for the experiments to be able to filter the truth
information, and indeed filter it differently than is done for a signal
event. What information is most of interest is both detector-specific
and analysis-specific, but frameworks must ensure the flexibility
required to satisfy even the most exotic use-case.

\hypertarget{physics-modeling}{%
\section{Physics Modeling}\label{physics-modeling}}

Geant4 physics libraries provide accurate simulation for a wide variety
of applications in HEP, space science, medical physics and other
scientific domains. HEP experiments tune their simulations by comparing
simulation results with both test beam data and collision data. By and
large, these comparisons show good agreement. Nevertheless,
discrepancies in comparisons between the results of simulation and
collider data do exist, and it can be extremely difficult to disentangle
whether the differences arise from misrepresentation of the detector
geometry and material or from inaccuracies in the physics modeling.

Big efforts continue to be invested by experimentalists to understand
the performance of their detectors, as well as by the physics developers
to improve and validate their physics simulation models. The comparison
against thin-target data has been very successfully used to develop and
to initially tune the physics models. However, it has been observed
recently that further tuning leading to improvements against thin target
data, does not necessarily lead to better simulation results when
compared to the thick-target data. Beyond the lack of enough thin target
data, this situation is, most likely, showing that the current models
are reaching their limits and in order to improve the overall detector
simulation, one needs to go beyond simple tuning of the existing models
and to develop new, more sophisticated ones. Another modeling challenge
that we will face at higher energies is related to the extreme closeness
of the central trackers to the interaction points, allowing short lived
particles such as hyperons to interact with the detector. Up to now,
these particles are decayed by the generators without being transported
by the simulation code as their decay happens in the vacuum of the beam
pipe. As beam energies increase, the line between particles dealt with
by generators and Geant4 will blur. These will survive long enough to
pass through the first few layers of the detectors leading to "missing
hits" in MC. As this situation becomes more common, the range of physics
models available for such particles will need to be expanded.

Our goal in detector simulation in the coming decade is to cope with
experimental needs for analysis of high-statistics data and for modeling
the next generation detectors (ILC/CLIC, FCC). Higher statistics in
future experiments will require smaller simulation errors, which in turn
implies higher precision models and larger Monte Carlo (MC) event
samples to be generated. This imposes two major challenges on physics
modeling. The first priority is to improve physics models to limit the
systematic error and match higher precision crucial for high luminosity
HL-LHC. The second priority is the extension of the validity of the
physics models up to FCC energies.

In the following, we summarise recent developments in electromagnetic
and hadronic physics modeling. We discuss physics accuracy and
performance improvements along with results of comparison of simulation
with experimental data.

\hypertarget{electromagnetic-physics}{%
\subsection{Electromagnetic Physics}\label{electromagnetic-physics}}

Electromagnetic (EM) transport simulation is challenging as it occupies
a significant part of the computing resources used in full detector
simulation. More experimental data in the coming decade implies higher
statistics and this implies the need for large and accurate simulation
samples, which require both model accuracy and speed. Therefore the
physics models need to be improved for achieving higher accuracy and the
implementation of algorithms needs to be reviewed in order to optimise
performance. Significant efforts have been made in the recent past to
better describe the simulation of electromagnetic shower shapes, in
particular to model the H $\rightarrow \gamma \gamma $ 
signal accurately at the LHC. This effort is being continued with 
emphasis on reviewing the
models assumptions, approximations and limitations, especially at very
high energy, and with a view to improving their respective software
implementations. For LHC Run-1 simulation Geant4 version 9.4 was used
~\cite{1742-6596-331-3-032029}. For Run-2 significant improvements were introduced into the
bremsstrahlung, pair production and multiple scattering,
further improving simulation of the EM shower shape 
~\cite{1742-6596-396-2-022013}~\cite{1742-6596-513-2-022015}; these improvements
were made available in Geant4 9.5 and 9.6. For validation purposes, an
extended set of tests versus data has been developed for high, moderate
and low-energies using test-beam data and published thin target data.
Comparisons with low-energy EM models from the Livermore and Penelope
packages were also carried out.

The process of multiple scattering of charged particles is a key
component of Monte Carlo transport codes. At high energy, it defines the
deviation of charged particles from ideal tracks, limiting the spatial
resolution of detectors. The scattering of low energy electrons defines
the energy flow across volume boundaries. This affects the sharing of
energy between absorbers and sensitive elements, directly affecting
shower shapes. A comprehensive description of recent improvements of the
Geant4 electromagnetic module can be found in~\cite{1742-6596-513-2-022015}. Rather good
agreement was found when Geant4 predictions were compared with
experimental data collected at the LHC.

Figure 2 shows a comparison of CMS test beam data with predictions from
two Geant4 physics lists, QGSP\_FTFP\_BERT\_EML and FTFP\_BERT\_EMM. EML
utilises a simplified (faster) multiple scattering model for all
detectors, while EMM uses the detailed multiple scattering model for
sampling calorimeters and the simplified model for other detectors. The
same figure also shows the ratio of Monte Carlo to data as a function of
beam momentum. 

\begin{figure}[bthp]
\centering
\includegraphics[width=0.94\textwidth]{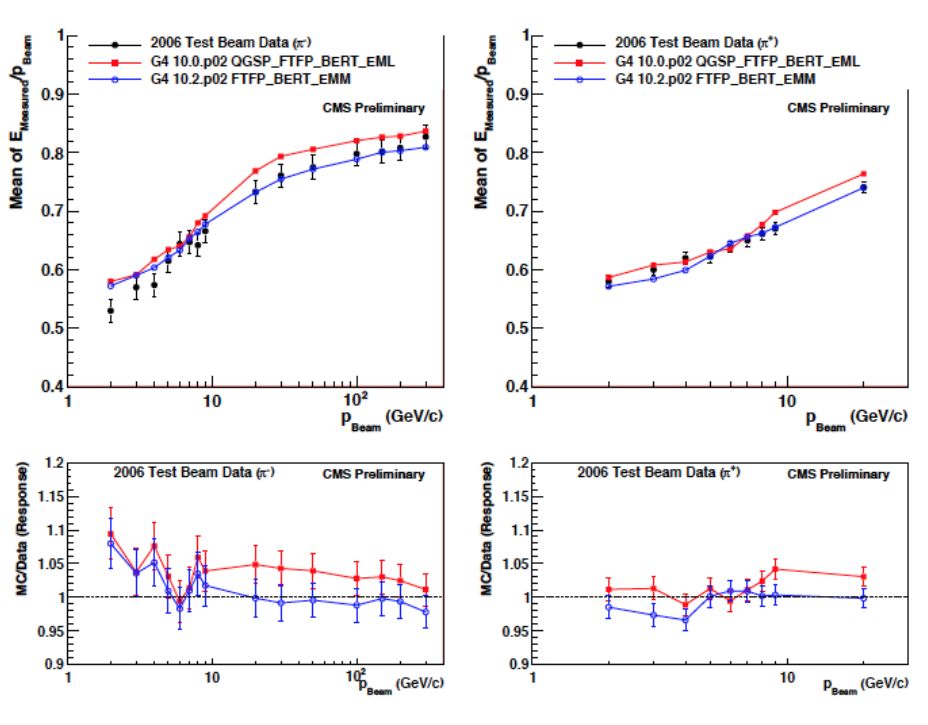}
\caption{The top plots correspond to the mean energy response for
p\textsuperscript{-} (left) and p\textsuperscript{+} (right) as a
function of the beam momentum. The bottom plots are ratios of Monte
Carlo predictions to the data for pions as a function of beam momentum.
The black points refer to the data, whilst the red and blue points are
Monte Carlo predictions from physics lists QGSP\_FTFP\_BERT\_EML of
Geant4 10.0.p02 and FTFP\_BERT\_EMM of Geant4 10.2.p02 respectively.}
\label{fig:enresp}
\end{figure}

It is clear from the figure that the predictions from the
physics list FTFP\_BERT\_EMM are closer to the data, in particular at
higher beam energies, and therefore CMS has taken the decision to move
to this list for its 2017 MC productions. Recent results from the CMS
experiment were also presented showing discrepancies in the EM shower
width in the endcap region as shown in Figure 3. This is a fresh
observation that needs a much more detailed understanding, leaving us
with an open question as to whether there is a problem in the physics
model or an issue with the modelling of detector materials.

\begin{figure}[bthp]
\centering
\includegraphics[width=0.94\textwidth]{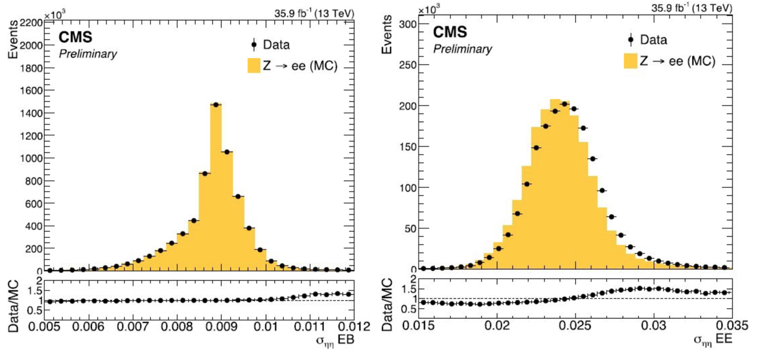}
\caption{Crystal-based EM shower width from the CMS experiment in the 
eta direction in both the barrel and the endcap. Simulation and data
are in good agreement in the barrel (figure on the left) but not in
the endcap region (figure on the right) where there is more material.}
\label{fig:emshwid}
\end{figure}

\begin{figure}[bthp]
\centering
\includegraphics[width=0.94\textwidth]{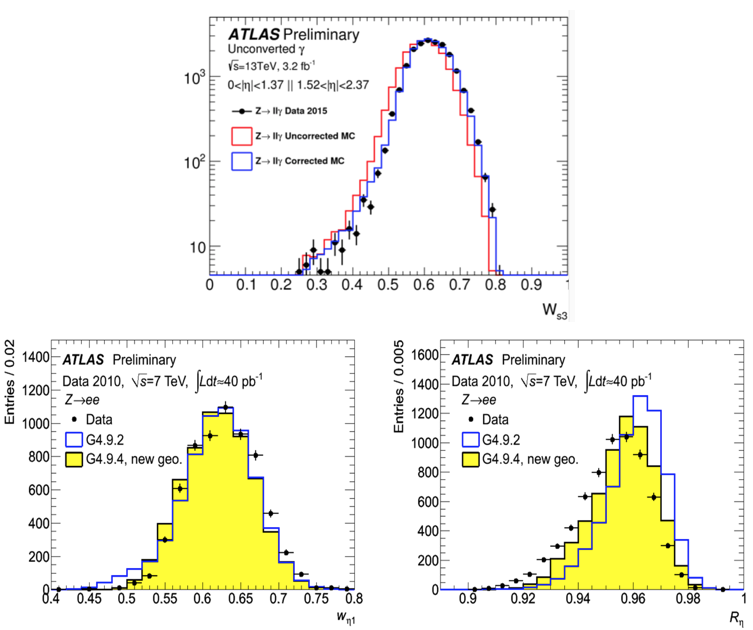}
\caption{The upper plot plot shows a measurement of R\textsubscript{$\eta$}
for unconverted photons in the ATLAS EM calorimeter, where
R\textsubscript{$\eta$} is the ratio of shower energy deposited in 3x7
cells to 7x7 cells in $\eta$-$\varphi$. The lower plot shows the lateral shower
width, W\textsubscript{$\eta1$}, for electrons in the first layer of the EM
calorimeter. Both plots show a clear discrepancy between real data and
the results of MC simulation.}
\label{fig:R}
\end{figure}

In the ATLAS experiment one area of disagreement concerns the
measurement of lateral shower shapes in the EM calorimeter. Figure 4
shows a measurement of R\textsubscript{$\eta$}  for unconverted photons in the
ATLAS EM calorimeter, where R\textsubscript{$\eta$} 
is the ratio of shower
energy deposited in 3x7 cells to 7x7 cells in the forward hadron
calorimeter. There is a clear discrepancy between real data and the
results of MC simulation with Geant4 9.6 (red line). Shower shapes in
eta-direction are consistently wider in data than simulation, both for
electrons and photons. Detailed investigations suggest that this is not
a material issue and therefore work is continuing to better understand
the impact of the EM physics modeling on the results of the simulation.
A similar discrepancy is seen in the measurement of the lateral shower
width for electrons in the first layer of the EM calorimeter, as can be
seen from the lower plots in Figure 4 obtained with Geant4 9.2 and 9.4.
Although the simulation in the upper plot can be adjusted to better
match the data (``corrected'', in the figure), it is challenging to
understand correlations between mis-modeled effects, and a more accurate
simulation is desirable. A substantial amount of work and high level of
coordination with experiments is therefore needed to investigate the
pending discrepancies between data and simulation.

The HL-LHC era will require higher precision simulations that can only
be achieved by continuous improvement of the electromagnetic physics
models used in the simulation. The further development of these models
is therefore still being actively pursued in order to improve their
precision and to extend their validity to higher energies. They include
the following.

\begin{itemize}
\item
  A new Goudsmit-Saunderson ``theory-based'' model for describing the
  \emph{multiple scattering} of electrons and positrons has been
  developed (Figure 5, blue line). It has been demonstrated to
  outperform, in terms of physics accuracy, the old theory-based model
  in Geant4 (grey line) and the Urban model that is the Geant4 default
  (Figure 5). The new implementation incorporates all the ingredients of
  the theoretical model resulting in extremely smooth angular
  distributions and has been developed with a view to fully exploiting
  track-level vectorisation in the future. At the same time it is
  important to note that this was achieved without incurring a CPU
  performance penalty.
\end{itemize}

\begin{figure}[bthp]
\centering
\includegraphics[width=0.94\textwidth]{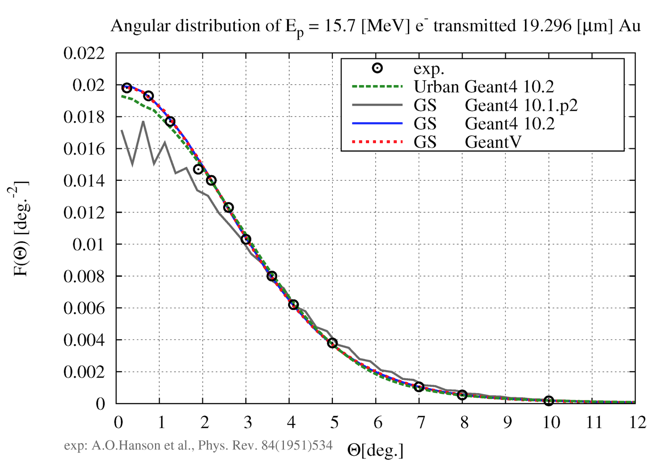}
\caption{The new Goudsmit-Saunderson multiple scattering
model implementation (GS - Geant4 10.2) shows no artefacts and excellent
agreement with experimental data}
\label{fig:gs}
\end{figure}

\begin{itemize}
\item
  The models used to describe the \emph{bremsstrahlung} process have
  also been reviewed for Geant4 9.5 allowing to describe the shower
  shape for the CMS barrel (Figure 3). Recently an improved theoretical
  description of the Landau-Pomeranchuk-Migdal (LPM) effect was
  introduced which plays a significant role at high energies. We show in
  Figure 6 a comparison between the old (grey line) and improved
  bremsstrahlung implementations with respect to experimental data. The
  review of this model was beneficial and demonstrated excellent
  agreement with the data. Once again, these improvements were made
  without introducing any speed penalty. They are available with Geant4
  10.3.
\item
  The model describing the pair-production process has also been
  reviewed. Similar improvements were introduced in the LPM suppression
  functions. At very high energy the improved pair production model
  shows a significant difference in the differential cross-section. The
  old implementation (green plot in Figure 7) was over-suppressing $ \gamma$
  conversion i.e. the LPM effect was overestimated. There are currently
  no available experimental data for validation of our results. The use
  of the relativistic pair-production model is essential for accurate
  physics simulations for FCC.
\item
  At extreme high energies (important for FCC studies), nuclear size
  effects become important. Different elastic scattering models with
  nuclear form factors were developed taking into account finite nuclear
  size effects. Figure 8 shows that the newly developed single
  scattering model agrees with experimental data at relatively low
  energy. Also at high energies, since precision becomes increasingly
  important, formfactor description may affect various EM processes.
  Such effects may be important for high energy muons, for shower shape,
  and for simulation of background for a dark matter search experiment.
\end{itemize}

\begin{figure}[bthp]
\centering
\includegraphics[width=0.94\textwidth]{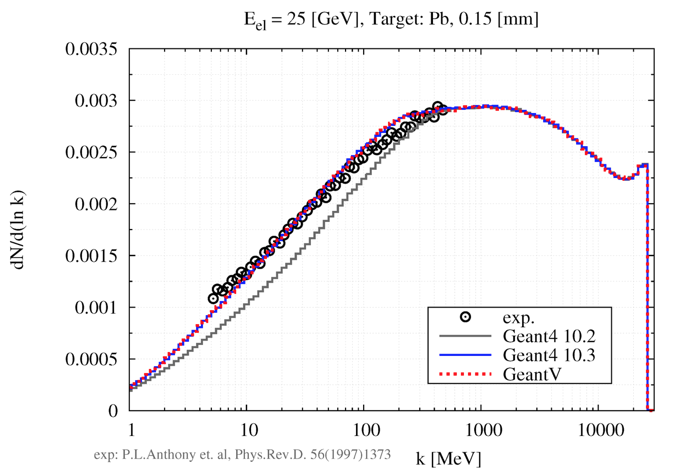}
\caption{The improved bremsstrahlung model included in Geant4
10.3.beta and in GeantV shows better agreement with experimental data
with respect to the existing implementation}
\label{fig:brem}
\end{figure}

\begin{figure}[bthp]
\centering
\includegraphics[width=0.94\textwidth]{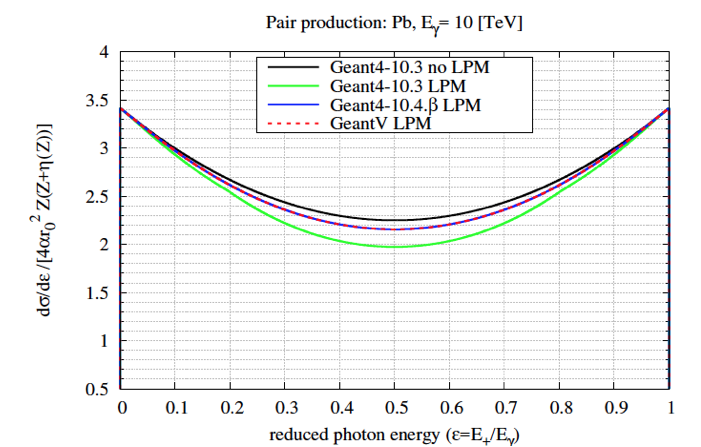}
\caption{The improved pair production model included in both
Geant4 10.4.beta and in GeantV is compared with the existing Geant4
implementation with and without LPM}
\label{fig:ppm}
\end{figure}

\begin{figure}[bthp]
\centering
\includegraphics[width=0.94\textwidth]{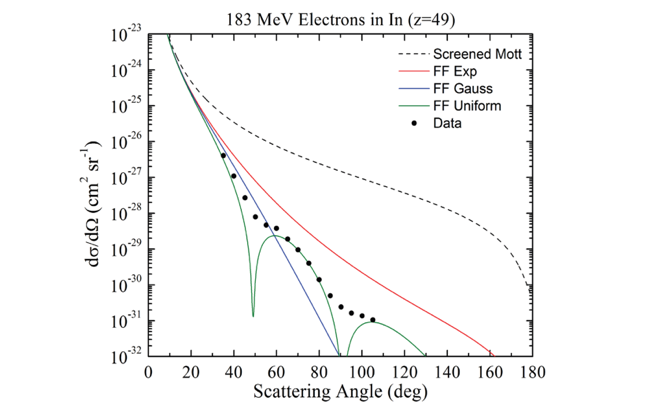}
\caption{New models were developed in Geant4 taking into account
finite nuclear size effect for elastic scattering of electrons at high
energies}
\label{fig:newm}
\end{figure}

Theoretical review of all electromagnetic models, including those of
hadrons and ions is therefore of high priority both for HL-LHC and even
more importantly for FCC studies. Efforts in this direction will
continue, but completion of this task by 2025 will demand a substantial
amount of work and continuous investments in effort. Avenues that should
be investigated include other channels that become important at extreme
energies, such as triplet production taking into account nuclear recoil
effects, size effects as well as $\gamma$ conversion to muon and hadron pairs.
Energy fluctuation models should be developed based on theory, not just
parametrisation. For Geant4 10.3 the upper energy limit for all EM
physics models was extended to 100 TeV, which is important not only the
HL-LHC programme but becomes crucial for the work being carried out in
the context of the FCC detector design studies. At the same time,
further improvements for high energy EM models are possible. A
substantial amount of work is needed for preparing simulations of next
generation detectors.

\hypertarget{hadronic-physics}{%
\subsection{Hadronic Physics}\label{hadronic-physics}}

Hadronic physics simulation is loosely defined to cover any reaction
producing hadrons in its final state. As such, it covers purely hadronic
interactions, lepton- and gamma-induced nuclear reactions, and
radioactive decay. In Geant4 the interaction is represented as a process
which consists of a cross-section to determine when the interaction will
occur, and a model which determines the final state of the interaction.
Models and cross-sections are provided which span an energy range from
sub-eV to TeV. More than one model is usually offered in any given
energy range in order to provide alternative approaches for different
applications. During the last several years, new models and
cross-sections have been added to the toolkit, while others have been
improved and some obsolete models have been removed.

it is not possible for a single model to describe all the physics
encountered in a simulation due to the large energy range that needs to
be covered and the simplified approximations that are used in hadronic
models to overcome the difficulty of solving the full theory (QCD). A
typical sequence of reactions may begin with a high energy
hadron--nucleon collision within a nucleus (parton string model),
followed by the propagation of the secondaries of the collision through
the nuclear medium (intra-nuclear cascade model), followed by the
de-excitation of the remnant nucleus (pre-compound model) and its
evaporation of particles (evaporation and breakup models) until it
reaches the ground state. Each of these stages is qualitatively
different from the other. Wherever possible a theory-based approach to
their implementation is followed as opposed to a phenomenological or
parameterised approach. The main motivation is to have more confidence
in the behaviour of the model outside the kinematic regions where it is
validated and tuned with experimental data, and also to preserve the
correlations between the different emitted particles.

Two models based on quark-parton concepts have been implemented in
Geant4, the quark--gluon string (QGS) model and the Fritiof (FTF) model.
There are also three intra-nuclear cascade models that are now offered:
Bertini, Binary and INCL++. As of release 10.0, the toolkit provides
nine reference physics lists whose names reflect the combination of
models used to describe the hadronic interactions necessary for various
applications over the whole energy range. Reference physics lists are
extensively and routinely validated. Currently the most-used reference
physics list for high energy and space applications is FTFP\_BERT. It
uses the Geant4 Bertini cascade for hadron--nucleus interactions from 0
to 5 GeV incident hadron energy, and the FTF parton string model for
hadron--nucleus interactions from 4 GeV upwards. The letter P indicates
that the Geant4 pre-compound model is used to de-excite the nucleus
after the high energy FTF interaction has been completed. The
FTFP--Bertini combination forms the backbone of many physics lists. It
successfully extends the use of the string model at lower energies, down
to 3 GeV, and at the same time the Bertini intra-nuclear cascade model
has been validated up to 15 GeV. The net result has been that the need
for using the LEP parameterised model at mid-range energies (around 10
GeV) has been eliminated. QGSP\_BERT is a popular alternative which
replaces the FTF model with the QGS model over the high energy range. It
is important to note that the existence of more than one model (for each
energy range) is very valuable in order to be able to determine the
systematics effects related to the approximations used. The situation is
similar to that of the Monte Carlo generators where the usage of
different generators provides an assessment of the systematic errors.

Detector response is an effective test of any model combination. It is
defined as the ratio of deposited energy visible to the detector, to the
incident beam energy. For the above combination of models (as in the
FTFP\_BERT physics list), the general agreement between the simulated
response and data for hadron-induced showers is at the level of a few
percent. Other useful data, such as shower shapes and energy resolution
are less precisely described and show agreement at a level of 10--20\%.

The use of highly granular calorimeters such as the ones being designed
by the CALICE collaboration for future linear colliders, allows a
detailed validation of the development of hadronic showers with
test-beam data. Preliminary results suggest that the lateral profiles of
Geant4 hadronic showers are too narrow. Comparisons with LHC test-beam
data have shown that a fundamental ingredient for improving the
description of the lateral development of showers is the use of
intermediate and low energy models that can describe the cascading of
hadrons in nuclear matter and the subsequent de-excitation of the
``wounded'' nucleus. The longitudinal development of hadron showers
mainly depends on the hadronic interactions at higher energies in the
forward direction: quasi-elastic scattering and diffraction. An
important effect recently introduced in Geant4 is the improvement of the
neutron capture cross-sections and final state generator. Based on the
high precision neutron library, it allows for an improved simulation of
the time structure and the lateral profile of hadronic showers in
neutron-rich materials ~\cite{WERNER200881}. Other recent developments include an
improved and re-tuned Fritiof model.

The LHC experiments are preparing detector upgrades for PHASE 2 of the
physics programme. For example, CMS will have a high granularity
calorimeter replacing the endcap electromagnetic and hadron
calorimeters. This will produce two main challenges. Firstly the EM
calorimeter will be of sampling type and so the longitudinal shower
profile becomes very important. Secondly, the sensitive detector
elements will be a mixture of silicon and scintillators and some of them
have hexagonal or half-hexagonal shape. This increases the complexity of
the detector geometry and will present a new challenge for the CPU
performance of the simulation code.

Work is ongoing to make a thorough review of the hadronics sector.
Additional work is currently being invested in the further improvement
of the QGS model, which is a more theory-based approach than the
phenomenological FTF model, and therefore offers better confidence at
high energies, up to a few TeV. This again is a large endeavour and
requires continuous effort over a long time. Further extension of the
coverage of hadronic interactions to higher energies, in the multi-TeV
region, will be necessary for FCC detector simulation studies. Moreover,
this would open a completely new domain of applications: the possibility
to simulate very high energy cosmic ray showers in the Earth's
atmosphere. The EPOS hadronic generator ~{20{]} offers currently the
most accurate simulations in this domain, as a fresh C++ rewrite of the
model, appears as a very interesting possibility for the future of
hadronic simulations in high energy and cosmic ray physics.

For the intra-nuclear cascade models, while Bertini and Binary are
likely to remain stable, further development is expected for the INCL++
model, besides the recent extension to higher energies (up to about 20
GeV). At the level of physics lists, the combined use of all the three
intra-nuclear cascade models - choosing between them according to
projectile hadron, projectile kinetic energy and target nucleus - will
be explored to optimise physics accuracy or simulation speed.

Low-energy hadronic physics - pre-equilibrium, nuclear de-excitation and
radiative decay models, as well as high-precision, data-driven
transportation of low-energy neutrons and charged particles (deuteron,
triton, He3 and alpha) - is and will remain a particular active area of
development, mostly driven by medical and nuclear physics applications.

The way hadronic cross-sections are implemented is critical for the CPU
performance of hadronic physics simulations and there are on-going
efforts to review them and make them more efficient via the development
of a common GeantV / Geant4 module. Total, inelastic and elastic
cross-sections for hadron--nucleus, nucleus--nucleus and
antinucleus--nucleus reactions are provided which cover energies up to
TeV in some cases. Proton-, neutron- and pion-nucleus cross-sections at
low to medium energies have been available in Geant4 since its
beginning. Recent work has focused on development of cross-sections for
other projectiles and at higher energies. The general behaviour of the
optical models is to predict constant cross-sections for very high
energies. However, experimental data show a moderate relativistic rise
of hadron--nucleus cross-sections. For this reason the Glauber model is
being used to describe hadron--nucleus cross-sections in the high energy
region (above 90 GeV). The current implementation of this model is under
review with a view to re-implementing it.

The above, ambitious program of development for the simulation of
hadronic physics relies heavily on an updated, extended, robust and
automated validation test-suite. Plenty of work has been done in the
recent past in this direction, but a lot more remains to be done. In
particular, a well-defined procedure to tune hadronic models needs to be
defined and deployed.

\hypertarget{physics-modelling-for-the-intensity-frontier}{%
\subsection{Physics Modelling for the Intensity
Frontier}\label{physics-modelling-for-the-intensity-frontier}}

Much of the text above concentrates on physics processes relevant to the
LHC experiments. Much of this overlaps the needs of muon and neutrino
experiments as well as dark matter and rare-decay experiments, which
have typically used Geant4 physics lists tuned based on LHC
requirements. However, as the physics programs of these experiments evolve,
they require simulation of increasing accuracy, which depends significantly
on the experiment but in all cases involves accurate modeling of hadrons and
electromagnetically interacting particles from a few MeV to a few GeV.
Precise neutron simulation is
particularly critical for accelerator-based neutrino experiments that
must reconstruct neutrino energy from final state particles. Also a
radioactive decay model for modelling the decay chain of isotopes is
essential for background studies.

Major components of the intensity frontier over the next decade will
include liquid Argon TPC detectors and liquid Xenon calorimeters.
For example, liquid Argon TPC's are equipped with optical readout systems 
to detect optical photons from the scintillation process. The information 
from the optical system is used to  provide a trigger and, potentially, 
could be used to improve the energy resolution of showers by combining the 
scintillation and charge collection signals. The modeling of optical photons 
in liquid Argon detectors is a real challenge, since 50,000 scintillation 
photons are produced per 1 MeV of energy deposited in liquid Argon. To track 
all these photons in Geant4 would increase the computing time on a CPU by a 
factor of  more than one thousand. Therefore, the full simulation is only used 
to produce lookup tables with the probability that that an optical photon makes 
it to the photo detector. These tables are an approximation and cannot replace 
the full simulation without a significant loss in physics accuracy. Even the 
lookup table solution requires a significant amount of CPU time to produce 
enough statistics. In order to satisfy the physics accuracy and computing 
performance requirements of future neutrino experiments, approaches based 
on new computing platforms are under investigation. For example, 
Opticks ~\cite{1742-6596-898-4-042001} 
is a software package where the optical physics processes are implemented on the GPU. 
This package is executed on NVIDIA machines to parallelize the
tracing of optical photons and yields a drastic time performance improvement 
with a speedup of the order of a thousand with respect to the original Geant4 
package running on CPUs. Opticks has been used to simulate the optical photons 
in the JUNO neutrino experiment. 

Efforts to address common needs of the liquid Argon experiments
including the post Geant4 phase of simulation are undertaken in projects
such as LArSoft~\cite{LArSoft} and Qscan~\cite{LArLEM-TPC}, while extending Geant4 EM
physics models for liquid noble gases is undertaken by the NEST project
~\cite{Szydagis:2013sih}. Detectors with large numbers of wires and photo-sensors may
naturally lend themselves to parallelisation techniques, provided that
the simulation frameworks are able to handle such cases in an efficient
way. Additionally, the Wire Cell Toolkit ~\cite{Wire-Cell} provides simulation,
noise filtering, signal processing and reconstruction algorithms for
LArTPC detectors. It includes experimental support for parallel
processing following the data flow processing paradigm. More work and
input from parallel processing experts is needed to bring this support
to maturity.

Data-taking at current experiments, such as MicroBooNE, will greatly
inform the simulation needs of the Intensity Frontier (IF) community
moving forward. This might be most relevant for LArTPC technology as it
is the newest and least familiar. For example, if significant
non-uniformity of purity in data is observed in MicroBooNE, simulation
of non-uniform purity will be a high priority simulation need of future
LArTPC experiments.

Efforts are also underway to improve the accuracy of Geant4. While much
of this effort was motivated by requests from Intensity Frontier
experiments, especially the variation of the model parameters, it will
be of use to the entire HEP community. This work includes:

\begin{itemize}
\item
  Validation of models: a validation database known as DoSSiER ~\cite{Wenzel:2017dqo}
  that contains data from experiments measuring particle cross-sections
  (e.g. NA61) is being developed. It was started as a Geant4 project,
  but is now also used by GeantV and should be suitable for other MC
  toolkits such as e.g. GENIE.
\item
  Development of new physics lists: Geant4 physics lists utilised by IF
  experiments were originally designed to meet the needs of LHC
  experiments. For example, a ``Shielding'' physics list was originally
  configured for radiation shielding studies and cavern background
  simulations for LHC experiments, but it is now widely used by
  underground dark matter experiments such as CDMS and LZ. A variant of
  this Shielding physics list (ShieldingM) was created to address the
  needs of Mu2e. Geant4 physics lists tailored for LHC were not adequate
  to the generation of neutrino beams, they disagree with existing
  hadron-production data by up to 40\% in some areas of phase space. In
  light of this, a new physics list (``NuBeam'') has been created, aimed
  at meeting the specific needs of neutrino beam simulations.
\item
  Variation of model parameters: Equally as critical as having an
  accurate beam or detector simulation is having an ability to quantify
  that accuracy. Neutrino experiments go to great effort to estimate and
  propagate uncertainties on Geant4 models to physics measurements.This
  is currently done experiment-by-experiment, but an initiative within
  Geant4 was recently begun to provide methods of assessing model
  parameters. Additionally, procedures are being developed to modify
  model parameters and compare with data, with the aim of producing
  uncertainties and covariance matrices on these parameters that can
  then be propagated to physics measurements.
\end{itemize}

\hypertarget{non-hep-use-cases}{%
\subsection{Non-HEP use-cases}\label{non-hep-use-cases}}

The extensibility of Geant4 has been appreciated also outside of the HEP
domain and several applications and extensions have been created to
perform simulations in some additional domains. It is worthwhile to
mention:

\begin{itemize}
\item
  Crystal extensions to model the property of crystalline structures.
  Applications include the use of bent crystals for beams manipulations,
  phonons and ultra-cold neutrons beams ~\cite{Brandt:2014imy}.
\item
  Geant4-DNA: a Geant4 extension for the calculation of physico-chemical
  damage to DNA at a nanoscale level ~\cite{bernal:hal-01288764}
\end{itemize}

\hypertarget{radiation-and-non-collision-background-modeling}{%
\subsection{Radiation and Non-Collision Background
Modeling}\label{radiation-and-non-collision-background-modeling}}

The LHC experiments have made use of the FLUKA simulation package, and
to a lesser extent GCALOR and MARS, for modeling of the radiation
backgrounds, primarily of low-energy neutrons and photons, present
during and after the running of the LHC. Understanding these backgrounds
is critical for estimating the required radiation hardness of
on-detector electronics, for developing shielding that might reduce
non-collision backgrounds in the detector, and for evaluating radiation
safety procedures when servicing or upgrading the detectors. The FLUKA
simulation package provides excellent modeling of these backgrounds, but
to date has not been used for any significant high-energy collision
modeling by any of the experiments. While its maintenance and continued
functionality is critical for the future of the LHC experiments and for
the development of the next generation of experiments, improvements in
both performance and physics are not as critical as they are in the case
of Geant4.

\hypertarget{software-performance}{%
\section{Software Performance}\label{software-performance}}

Simulation, including physics generation, interaction with matter
(detailed analogue or parametrised simulation), digitisation,
reconstruction and analysis takes a large fraction of the computing
resources utilised by HEP experiments. Depending on the experiments, and
the complexity of the detectors, each of these components of the
simulation chain contribute with different weights to the total. Setting
aside the reconstruction of simulated events, the modeling of the
interaction with matter is the most expensive module of a simulation
application, as is certainly the case for the LHC experiments. Neutrino
detectors are typically simpler, with a larger fraction of resources
spent in the simulation of the readout. Physics generation usually takes
the smallest fraction, except when generating signal samples that
require a scan of large regions of theory parameter space or for cutting
edge, high-accuracy calculations that are now becoming possible with new
event generation tools.

As an illustration of the above, the CMS experiment, from start-up in
2009 through May 2016, the Full Simulation chain including all elements
described above took 85\% of the total CPU time utilised by the
experiment, while the Geant4 module took about 40\%. This information
was obtained from the CMS Dashboard, which is a computing information
monitoring source available to CMS members. The rest of the CPU cycles
were primarily used to reconstruct and analyze real collider data. The
assumption for the 85\% figure is that the analysis of simulated data
consumes 75\% of the CPU time spent in analysis, including both
simulated and real data, and excludes the generation of signal samples
for BSM searches. The reason why the analysis of simulated data takes a
larger fraction of the total analysis CPU time than the analysis of real
collider data is that the design and optimisation of the measurements,
as well as the development and validation of data-driven methods, are
all based on MC samples.

An analysis of jobs running on both the grid and on HPC systems in
ATLAS, between July 2015 and July 2016, showed similar results with
simulation utilising 65\% of the total time available to the experiment.
The breakdown shows that Event Generation (Evgen) took 25\%, thanks to a
big burst of Mira throughput, whilst Full Simulation (Geant4) took 39\%
and Fast Simulation (Atlfast-II) just 2\%. In addition, Event
Reconstruction of data and simulated events took 19\%, whereas the
production of derived datasets for use in analysis took 5\% and
user-level Analysis took 10\% .

The Geant4 module of the ALICE Full Simulation application takes 55\% of
the total detector simulation CPU time, excluding event generation and
reconstruction, while digitisation takes 35\% of the time.

Although neutrino experiments are similar to LHC experiments in the
fraction of computing resources used for simulation, NOvA spent 70\% of
their CPU cycles on simulation in 2016, the total amount of computing
resources utilised by this experiment is two orders of magnitude less
than those spent by CMS.

The main drivers of the need to improve computing performance in the
next two decades are therefore the LHC experiments. It is expected that
their computing needs will increase by a factor of 10 to 100 in the
High-Luminosity LHC (HL-LHC) era depending on the solutions developed to
face simulation, pileup, and reconstruction challenges arising from the
high-luminosity environment. The challenge is daunting and the need to
explore both evolutionary and breakthrough solutions through R\&D
programs to speed up simulation and reconstruction is essential. The
computing needs of neutrino and muon experiments is expected to be
smaller than that of the LHC experiments, but will grow substantially
with time as more experiments come on-line and existing experiments
accumulate data. These experiments will therefore need to make use of
the solutions developed for the LHC.

The Geant4 Collaboration has invested significant effort into improving
the toolkit computing performance in the past decade, as code was
reviewed and optimised. In 2013, the release of event-level
multithreading capabilities in Geant4 brought significant memory
savings, as illustrated in Figure 9, which shows the CPU time (top) and
memory (bottom) consumption of a simplified CMS standalone simulation
application, based on Geant4 version 10.3.p01, runs for 50 GeV pions.
Geant4 achieves almost perfect strong scaling (top plot) with the number
of available cores (57) and still gains about +30\% in hyperthreading
regime ~\cite{G4CPUPerf} ~\cite{Dotti:2016ors}. The large memory reduction obtained with
multi-threading is clearly visible (bottom). While a multi-process
approach can reach the O(10GB) memory consumption with a relatively
small number of threads (Geant4 version 9.6), it is only with a well
designed multi-threading approach, which shares the read-only data among
threads, that it is possible to run with the maximum number of threads
remaining in a memory budget of approximately 1 GB. Recent versions of
Geant4 have put emphasis on further reducing memory consumption to the
current level of about 10 MB needed for each additional thread.

Figure 10 shows the percentage change in CPU time performance taking
Geant4 version 10.0 as a reference, starting with Geant4 version 9.4.p02
(2010) and ending with Geant4 version 10.2 (2015) for the standalone CMS
application and a simple calorimeter configuration made of
Cu-Scintillator in a 4 Tesla magnetic field ~\cite{G4CPUPerf}.The study is
performed for 50 GeV e\textsuperscript{-}, p\textsuperscript{-}, and
protons, as well as for H $\rightarrow$ ZZ events. On average, the time performance
improvement through the life of the LHC experiments (2010-2015) is of
the order of 35\%. All tests were performed on the same hardware (AMC
Opteron 6128 HE @ 2 GHz) using the same compiler and operating system
(GCC4.9.2, Linux x86\_64). Remarkably, the percentage time performance
improvement during the period of time shown in the plots is in double
digits, even as the physics models were improved significantly for
accuracy, something that typically comes associated with a time
performance penalty. Although CPU profiling is a good starting point to
improve time performance, the biggest gains in Geant4 were achieved
while focusing on reducing the memory usage. Important speedup was
obtained as a direct result of both reducing the memory churn and
improving the physics algorithms. For example, the rewriting of the
Bertini physics model code yielded a reduced memory churn, making it
substantially faster.

Evolving compilers also offer opportunities for speed gains in
simulation applications. Figure 10 illustrates the impact of the use of
shared and static libraries in a simplified CMS application (HepExpMT
public benchmark) ~\cite{Farrell:2016ovs}. In this example, static linkage yields a
20\% gain, while Profile-Guided Optimisation (PGO) gives another 20\%.
Multithreaded applications benefit more from static linkage and PGO,
larger speedups as the number of threads increase. These options are
particularly interesting because they do not change the physics output
of the simulation. If this constraint can be relaxed additional speedups
can be obtained using additional Geant4 flags to speedup the simulation.

\begin{figure}[bthp]
\vspace*{0.3cm}
\centering
\includegraphics[width=0.94\textwidth]{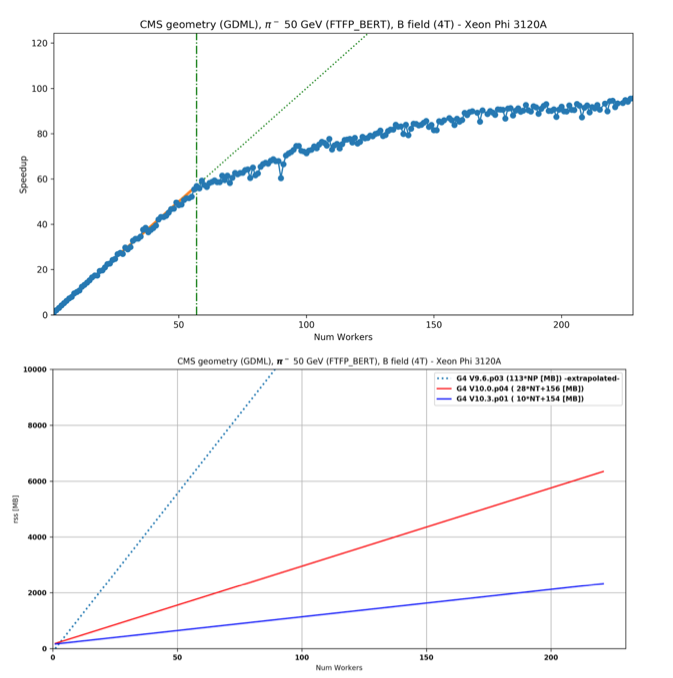}
\caption{CPU time (top) and memory (bottom) consumption of a CMS standalone
simulation application, based on Geant4 version 10.3.p01, run for 50 GeV
pions. The strong scaling (top) as a function of the number of threads
shows an almost perfect linearity (\textgreater{}90\%) with the number
of physical cores. The maximum number of cores (57) is denoted by the
dot-dash vertical line. The use of hyper-threading improves the
throughput by about 30\%. The use of multi-threading allows optimal use
of the limited memory budget of many-core systems. With multi-process
approaches (Geant4 9.6.p03 and 10.0.p04 sequential) it is possible to
use only a very limited set of the available parallelism. With
multithreading the memory requirements are reduced by a factor 10.}
\label{fig:perfcms}
\end{figure}

\begin{figure}[bthp]
\vspace*{0.3cm}
\centering
\includegraphics[width=0.94\textwidth]{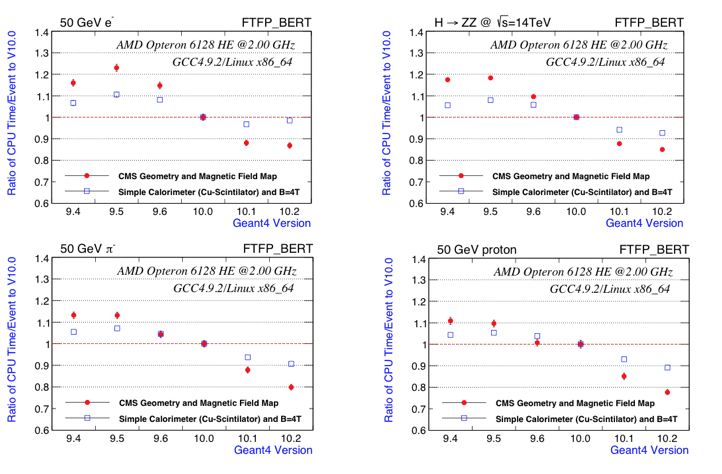}
\caption{Percentage change in CPU time performance taking Geant4
version 10.0 as a reference, starting with Geant4 version 9.4.p02 (2010)
and ending with Geant4 version 10.2 (2015) for the standalone CMS
application and a simple calorimeter configuration.}
\label{fig:perfchange}
\end{figure}

\begin{figure}[bthp]
\vspace*{0.3cm}
\centering
\includegraphics[width=0.94\textwidth]{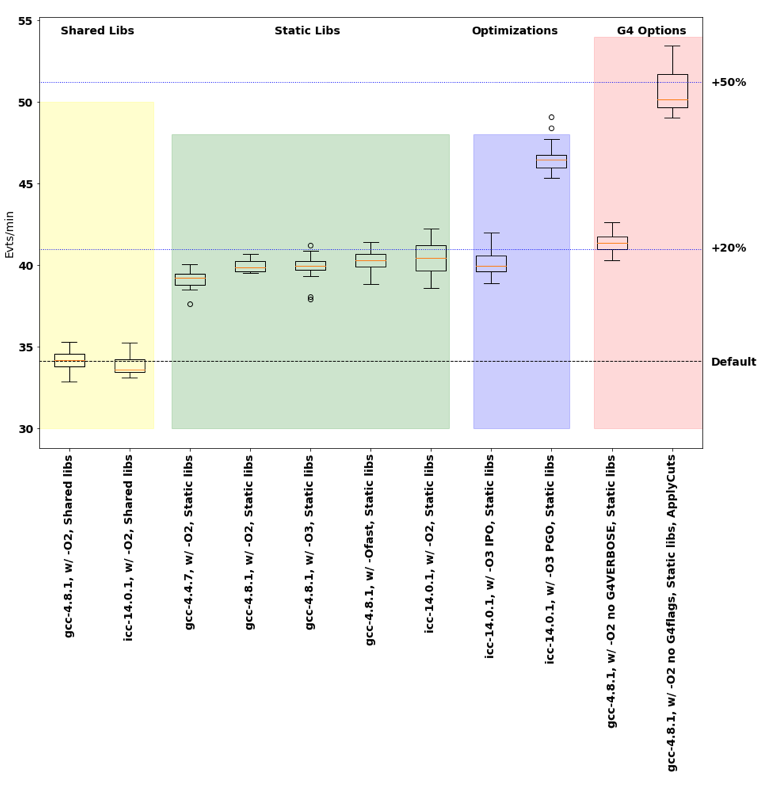}
\caption{Impact of the use of shared and static libraries in a
simplified HEP application. Results obtained with a simplified version
of CMS with Geant4 version 10.0. Performances for different combinations
of compilation options are compared. There is an important performance
boost when using static builds (+20\%), the main reason being the lack
of PLT calls. Another +20\% can be obtained using Profiler Guided
Optimisations (PGO). These techniques do not change the physics output
of the simulation, if this constraint can be relaxed compilation and
runtime options of Geant4 can be used to obtain another +10\% speedup. }
\label{fig:perflib}
\end{figure}

In the following sections, we will discuss computing performance studies
by different HEP experiments, as well as their diverse approaches to
improve computing performance.

\hypertarget{cpu-time-and-memory-performance-measurements}{%
\subsection{CPU Time and Memory Performance
Measurements}\label{cpu-time-and-memory-performance-measurements}}

The largest improvements to computing performance experienced by ATLAS
and CMS have come from physics code reviews as part of the effort to
reduce memory churn. The speedup has been a side effect of this effort.
The lesson learned is that profiling is not a very useful exercise
unless it is related to physics observables. The table in Figure 12
shows the number of Geant4 steps in each ATLAS sub-detector and for each
particle type, providing useful quantitative information to optimise
Geant4 tracking and therefore improve time performance. It is apparent
that most of the steps occur in the forward and EM calorimeters, for
electrons, photons, and neutrons. Based on the experience described
above, the Geant4 Collaboration is developing a tool to monitor the
number of steps and tracks, and the time spent in a given detector
element. The tool will also provide this information for each Geant4
track in user defined energy ranges. The output will be displayed in the
console, or saved in histograms or ntuples.

\begin{figure}[bthp]
\vspace*{0.3cm}
\centering
\includegraphics[width=0.94\textwidth]{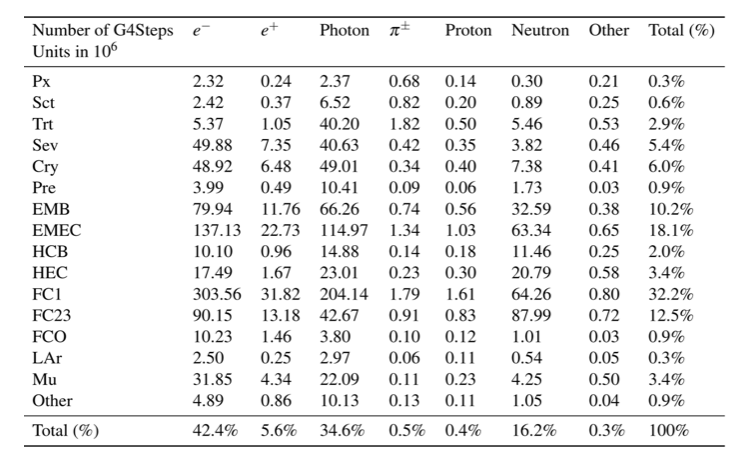}
\caption{Number of Geant4 steps in each ATLAS sub-detector and
for each particle type. The sub-detector types are: Px (pixel tracker),
Sct (strip tracker), Trt (straw-tube tracker), Sev (tracker services),
Cry (calorimeter cryostat), Pre (calorimeter presampler), EMB (EM barrel
calorimeter), EMEC (EM endcap calorimeter), HCB (hadronic barrel
calorimeter), HEC (hadronic endcap calorimeter), FC1 (forward
calorimeter EM module), FC23 (forward calorimeter hadronic modules), FCO
(other parts of the forward calorimeter), LAr (other parts of the liquid
argon calorimeter system), Mu (muon system), and other (those not
otherwise accounted for).}
\label{fig:perfsteps}
\end{figure}

Figure 13 shows a chart of CPU time for the G4 module of the CMS Full
Simulation application ~\cite{Hildreth:2017vpw}. About 60\% of the time is spent on
functions associated with the propagation of particles through the
detector geometry and magnetic field, while about 15\% is spent on EM
physics, and about 10\% followed by hadronic physics. Figure 14 shows
the CMS throughput and memory per event utilisation for QCD and ttbar
simulated events. Maximum CPU efficiency is achieved after
simulating about 500 events, when contribution of initialisation of
Geant4 and CMSSW become negligible.

\begin{figure}[bthp]
\vspace*{0.3cm}
\centering
\includegraphics[width=0.94\textwidth]{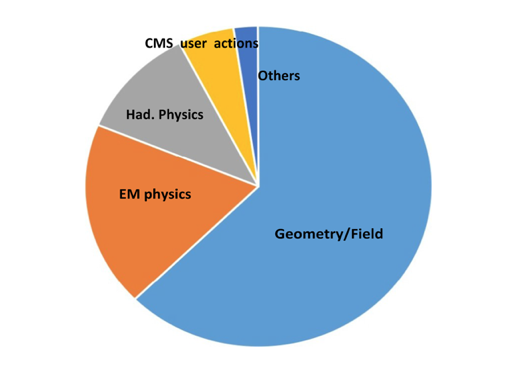}
\caption{Chart of CPU time spent in different tasks of the CMS
Full Simulation application for Run-2 simulation}
\label{fig:perfcpucms}
\end{figure}

Optimizing particle production cuts and replacing particle showers in
detectors with pre-generated showers or parameterisations are also part
of the strategy to improve computing performance. Both ATLAS and CMS
have these ``frozen showers'' or ``shower library'' options,
particularly to replace Full Simulation in the forward calorimeters.
GFLASH ~\cite{GRINDHAMMER1990469}, a Geant4 module for the fast simulation of
electromagnetic and hadronic showers using parameterisations for the
longitudinal and lateral profile, is not used at the moment due to the
peculiarities of the ATLAS accordion calorimeter and the limitations of
this approach to model longitudinal shower leakage in the CMS
calorimeters.

Geant4 offers the option to utilise the ``Russian Roulette'' technique
(see Sec. 10.1) to kill N-1 out of N particles within an ensemble of
neutrons and gamma below few MeV thresholds inside calorimeter geometry
regions. The surviving particle is assigned a weight of N. The use of
this technique by the CMS experiment yields significantly CPU time
savings, as discussed in Section 8.2.

\begin{figure}[bthp]
\vspace*{0.3cm}
\centering
\includegraphics[width=0.94\textwidth]{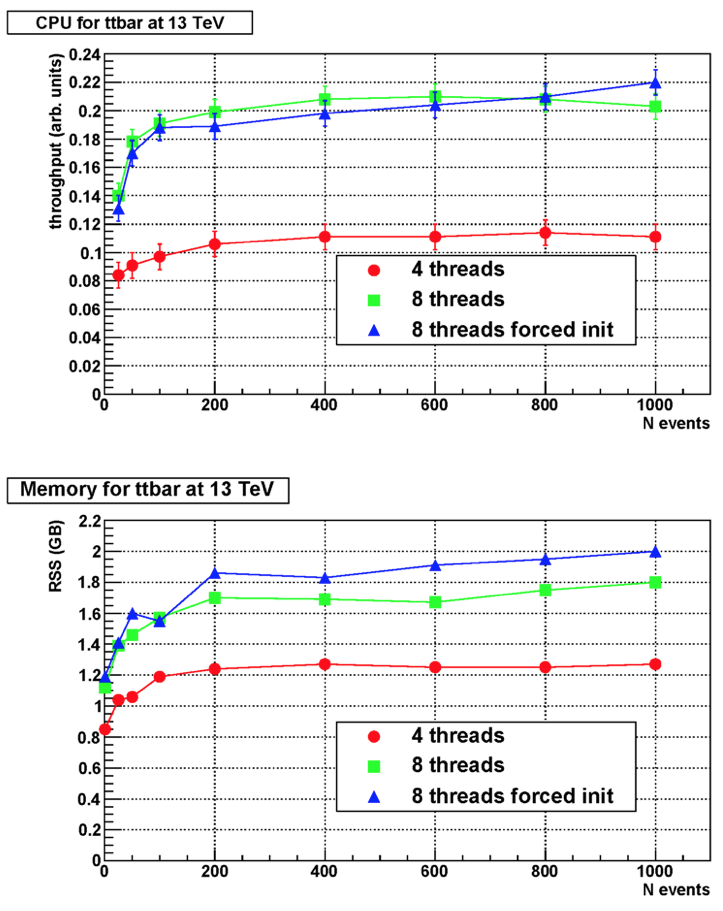}
\caption{Throughput and memory per event for ttbar simulated
events in CMS as a function of the event number. The version of Geant4
is 10.2p02, run in multi-threaded mode with 4 and 8 threads. Forced init
refers to the case when all data for photon evaporation is read at
initialisation time.}
\label{fig:perfthroughput}
\end{figure}

\hypertarget{ATLAS-and-CMS}{%
\subsection{ATLAS and CMS}\label{ATLAS-and-CMS}}

Given that ATLAS and CMS are expected to be the largest consumers of
computing resources a decade from now, the two experiments have decided
to develop a computing performance assessment program that includes the
Run 1 and Run 2 detectors as a reference, the different HL-LHC detector
scenarios to predict future needs, and an evaluation on how much can be
gained by utilizing state of the art techniques, such as re-optimizing
simulation parameters and replacing Geant4 showers with libraries of
frozen showers. Towards that end, ATLAS and CMS agreed on a set of
simulation configurations and physics input samples:

\begin{itemize}
\item
  Machine: olhswep16.cern.ch (CERN's OpenLab), one thread runs
\item
  Compiler: gcc 6.3
\item
  Geant4: version 10.2, FTFP\_BERT physics list
\item
  Pythia ~\cite{Sjostrand:2014zea} events: 13 TeV Pythia minimum bias (300 events) and
  ttbar (300 events), pseudo rapidity cut \textbar{}$\eta$\textbar{}\textless{}5.5
\item
  Particle gun: 50 GeV e's, muons, pions with a flat $\eta,\varphi$ distribution in
  $\eta={[}-0.8,0.8{]}, \eta={[}2,2.7{]}, and \varphi= {[}0,2\pi{]}$
\item
  Geometry: 2015 or 2016
\end{itemize}

From this reference configuration, various options are incorporated to
study the impact on time and memory performance, including Russian
Roulette, shower libraries, timing cuts, and particle production cuts
per region. For the reference configuration, each experiment uses its
own timing and production cuts, as well as its own choice of stepper and
tracking parameter values. The experiments also study the evolution of
the time and memory performance of the Full Simulation application as
the detectors are upgraded from the Run 1 and Run 2 configurations to
several potential HL-LHC configurations.

\begin{table}[bthp]
\vspace*{0.3cm}
\centering
\includegraphics[width=0.94\textwidth]{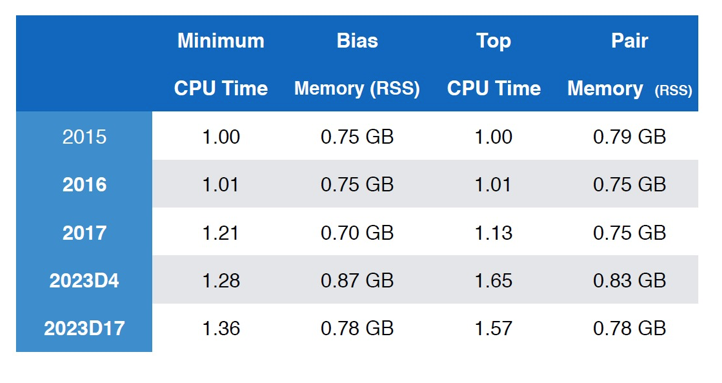}
\caption{Time and memory performance of the Geant4 module of the
CMS Full Simulation application for minimum bias and ttbar events in
current and future detector configurations. The Geant4 version used in
the test is 10.2p02. With respect to the 2015 and 2016 configurations,
which are very similar, 2017 adds an upgraded pixel detector and
modifies the forward calorimeter layer configuration, 2023D4 and 2023D17
are different versions of the HL-LHC CMS detector with an upgraded
tracker and a High Granularity end cap calorimeter (HGCAL), the latter
sub-detector involving a significant increase in the number of Geant4
volumes. }
\label{label:table1}
\end{table}

The differing structures of the ATLAS and CMS detectors means that even
in stand-alone Geant4 the simulation times are very different. Different
detectors imply different CPU demands. This is clearly demonstrated by
the information from CMS in Table 1. This shows the CPU time and memory
measurements for different CMS HL-LHC candidate detector configurations
with respect to the 2015 reference geometry. HL-LHC CMS configurations
demand 28-36\% more CPU cycles than 2015 or 2016 configurations to
simulate minimum bias events, while they require 57-65\% more for ttbar
events. Given that the physics of the simulation at the core of the CMS
HL-LHC upgrade, the High Granularity Calorimeter (HGCAL), is still not
fully validated awaiting further test beam experiments, new physics
accuracy requirements might push the CPU needs higher. The reason is
that the HGCAL is a significantly more complex apparatus than the
current detector using different technology and materials, covering the
{[}1.5,3{]} pseudorapidity range, consisting of 40 layers of silicon and
copper/tungsten, brass, or steel, with a total of 6 million channels
with cell sizes of 0.5 cm\textsuperscript{2} and 1
cm\textsuperscript{2}. Owing to much more modest detector revisions, the
difference in simulation performance for the current Run 2 and upgraded
ATLAS detector is 5\% for ttbar events and 15\% for minimum bias events.
This is simply a reflection of the geometry differences in the inner
detector and muon system, in the case of ATLAS, and no additional
penalty is expected from significant revisions to physics models.

Tables 2, 3 show the CMS time performance of the Geant4 module of the
Full Simulation application, that is excluding physics generation,
digitisation, and reconstruction, for the default configuration and
several options to reduce CPU time consumption in production of Pythia
generated minimum bias and ttbar events. Table 2 lists the
individual impact of each of the options, while Table 3 shows the
cumulative effect. The conclusion is that small gains may be achieved by
utilizing static libraries and by re-optimizing timing or particle
production cuts. 

\begin{table}[bthp]
\vspace*{0.3cm}
\centering
\includegraphics[width=0.94\textwidth]{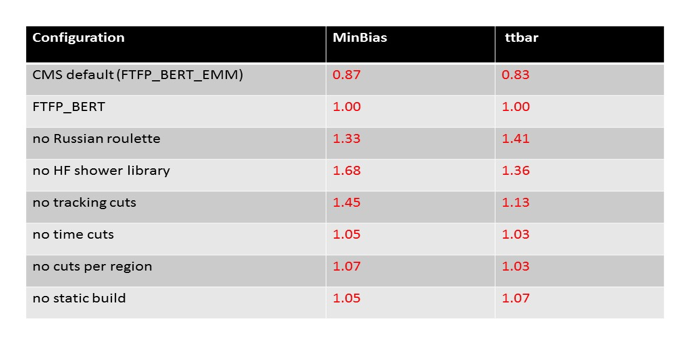}
\caption{Relative time performance of the Geant4 module of the CMS Full
Simulation application for minimum bias and ttbar production.
Measurements of the individual effect of several options to reduce CPU
time consumption are listed. The Geant4 version used in the test is
10.2.p02. Values of range cuts and tracking cuts are specified in the
text.}
\label{table:table2}
\end{table}

\begin{table}[bthp]
\vspace*{0.3cm}
\centering
\includegraphics[width=0.94\textwidth]{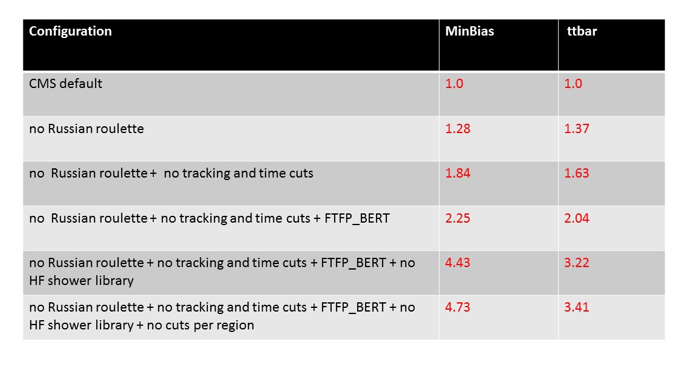}
\caption{Relative time performance of the Geant4 module of the CMS Full
Simulation application for minimum bias and ttbar production.
Measurements of the cumulative effect of several options to reduce CPU
time consumption are listed. The Geant4 version used in the test is
10.2p02. Values of range cuts and tracking cuts are specified in the text.}
\label{table:table3}
\end{table}

In the CMS simulation Geant4 time cuts and tracking
cuts are optimised per detector region. The cut in range value for the
production of secondary particles is 1 mm in the electromagnetic and
hadronic calorimeters, 0.01 mm in the pixel detector, 0.1 mm in the
strip tracker, 0.002 mm in the muon sensitive volumes, and 1 cm in the
support structures. The propagation time cut is 500 ns. The tracking cut
is applied for charged particles below 2 MeV inside the vacuum chamber
to avoid looping electrons or positrons. The Russian roulette technique
applied inside calorimeter regions gives a 33\% speedup and is already
utilised in current detector configurations for Run-2, as is also the
case of the HF shower library, which yielded a 68\% gain in time
performance. Additional CPU time savings are achieved by optimizing
parameters associated with particle tracking in EM fields and production
of secondary particles.

Tables 4 and 5 shows the relative CPU performance of a number of
configurations of the ATLAS simulation application. The largest gain
comes from introducing shower libraries in the forward calorimeter,
which comes at a cost of 430MB of VMEM per job. No other option applied
here has an appreciable effect on memory requirements of the
application. The neutron time cut shows only a small CPU improvement,
but was shown to reduce the size of the output file by almost a factor
of two in previous tests. This file size issue can also be mitigated by
various approaches to hit storage, however. Removing range cuts
decreases the simulation time per event, because the thin volumes of the
sampling calorimeter demand a tighter range cut, including 30~$\mu$m in the
forward calorimeter and 300~$\mu$m in the EM calorimeters, to provide an
accurate physics description. This penalty is not present in the CMS
simulation, since the CMS EM calorimeter has much larger active volumes.
Still, the effect of range cuts on the two applications is similar.

\begin{table}[bhp]
\centering
\begin{tabular}{| p{9cm} | c | c |}
\hline
Configuration & MinBias & ttbar \\ \hline
Nominal production configuration: shower libraries 
in the forward calorimeter, nominal range cuts, 
NystromRK4 stepper, FTFP\_BERT\_ATL
physics list, 250ns neutron time cut, simulation 
of primary particles with pseudo-rapidity below 6.0
 & 1.0 & 1.0 \\ \hline
No shower libraries & 1.5 & 1.3 \\ \hline
ClassicalRK4 stepper instead of NystromRK4 & 1.09 & 1.07 \\ \hline
No neutron time cut & 1.02 & 1.01 \\ \hline
FTFP\_BERT instead of FTFP\_BERT\_ATL physics list & No change & No
change \\ \hline
No simulation of primaries with pseudo-rapidity above 5.5 & 0.94 & 0.95
\\ \hline
All range cuts set to 1mm & 0.92 & 0.90 \\ \hline
\end{tabular}
\caption{Performance of various configurations of the ATLAS
simulation for minimum bias and ttbar production events. The Geant4
version used for this test was G4 10.2p03. No significant performance
improvements were introduced in patch 03 with respect to patch 02.}
\label{table:table4}
\end{table}

\begin{table}[bthp]
\centering
\begin{tabular}{| p{9cm} | c | c |}
\hline
Configuration & MinBias & ttbar \\ \hline
Nominal production configuration: & 1.0 & 1.0 \\ \hline
No shower libraries & 1.5 & 1.3 \\ \hline
No shower libraries + use ClassicalRK4 stepper & 1.6 & 1.4 \\ \hline
No shower libraries + use ClassicalRK4 stepper + No neutron time cut &
1.6 & 1.4 \\ \hline
No shower libraries + use ClassicalRK4 stepper + No neutron time cut +
use FTFP\_BERT physics list & 1.6 & 1.4 \\ \hline
No shower libraries + use ClassicalRK4 stepper + No neutron time cut +
use FTFP\_BERT physics list + All range cuts set to 1mm & 1.5 & 1.3
\\ \hline
No shower libraries + use ClassicalRK4 stepper + No neutron time cut +
use FTFP\_BERT physics list + All range cuts set to 1mm + No simulation
of primaries with pseudo-rapidity above 5.5 & 1.4 & 1.2 \\ \hline
\end{tabular}
\caption{Performance of various configurations of the ATLAS
simulation for minimum bias and ttbar production events. Measurements of
the cumulative effect of several options to reduce CPU time consumption
are listed. The Geant4 version used for this test was G4 10.2p03. No
significant performance improvements were introduced in patch 03 with
respect to patch 02.}
\label{table:table5}
\end{table}

Having exhausted the most obvious optimisations, CMS is exploring
various opportunities to improve computing performance in various ways
e.g. by testing the newest Geant4 versions, by using the most advanced
compilers, and by implementing an energy dependent ``smart tracking''
algorithm for propagation of particles in EM fields. However, the
biggest gains will most probably come from ongoing R\&D efforts, such as
the vectorised geometry library, VecGeom, which was recently released
for use with Geant4. Although VecGeom is optimised for GeantV, where it
promises a significant speedup by benefiting from data locality and SIMD
vectorisation, preliminary tests with Geant4 in CMS give a 5-10\% time
performance gain coming from improvements in the geometry algorithms. In
other words, although VecGeom can only be run in scalar mode with
Geant4, it still delivers significant time performance gains. The GeantV
project will offer more opportunities in the future, with the
possibility of integrating navigation and EM physics libraries to the
Geant4 CMS simulation application. The promise of a factor of 2-5 time
performance improvement using GeantV's full track-level parallelisation
approach and SIMD vectorisation is very enticing, and plans are being
made for a potential migration within the timeframe of the start of the
HL-LHC run in 2026, if the GeantV computing performance targets can be
met.

ATLAS is also exploring a number of technical and physics options to
improve simulation performance. Several of these options were delayed in
order to complete a migration to a new multi-threading friendly
infrastructure and to use a cmake build system. Russian Roulette for
neutrons and static library building are both currently being explored,
and it is expected that PGO will be investigated after those are
complete. Some further optimisation of range cuts is to be validated in
order to gain a few percent of CPU performance. ATLAS has also
investigated the new geometry modules available in the Geant4 10.4 beta
release and found in some cases significant CPU improvements resulting
from these (5\%-level). They are expected to be validated and moved into
production soon. VecGeom will be investigated as well.

When considering which optimisations to implement, or even the order in
which they should be attempted, the total amount of effort required to
bring each one into production should be taken into account, as well as
the overall impact on resource requirements.

An assessment of the effort required, in addition to the obvious cost of
how much person-power is required to implement an optimisation, should
also include how complicated it is to validate. For example static
compilation of Geant4 should come with no change in physics, so it is
easy, in principle, to validate. Conversely changing range cuts will
very often change the physics and so will require a lot more careful
examination by a larger group of experts before it can be signed off. A
modest CPU improvement may not be worth the risk of reproducing a large
Monte Carlo dataset if a problem is found with the physics of the
simulation.

When making an assessment of the overall impact on resources, the
potential gains in CPU usage and potential reductions in disk usage
should be considered alongside other potential issues. For example, if
samples simulated with the new optimisation cannot be combined with
older samples, then a new simulation campaign with new calibrations
might be required to benefit from the new optimisation.

\hypertarget{lhcb-and-alice}{%
\subsection{LHCb and Alice}\label{lhcb-and-alice}}

Although the use of CPU grid resources by LHCb is two orders of
magnitude below ATLAS', the expectation for the 2020 LHC run and beyond
is to use up to 90\% of its allocation for Monte Carlo simulation. In
Run2 the experiment has adopted the concepts of split High Level
Triggers (HLT) and of the TURBO stream that integrate real-time
calibration and alignment into the data taking process, thereby
performing an online reconstruction that is equivalent to offline.
Nevertheless LHCb is currently exploiting all resources at its disposal
for simulation, including the use of its Online Farm when idle. The full
readout of the detector at 40 MHz in the LHCb upgrade will have a major
impact on trigger and offline computing systems and the experiment will
have to cope with higher rates, up to 2-5 GB/s from the detector. Hence
the concepts above will be further exploited moving the event
reconstruction and selection as close as possible to the online farm.
The experiment plans to match the amount of physics events collected
with the same order of simulated events. Exploring faster simulation
options is therefore a primary focus of the LHCb software efforts. The
experiment has yet to evolve its framework for parallelisation and is
currently migrating to a version of Geant4 that supports multithreading,
with the promise to achieve significant memory savings. Potential gains
from static library linkage and re-optimisation of simulation parameters
will also be investigated. The plan is to evolve the software into an
integrated simulation framework which allows to choose and combine
different fast simulation options. Table 6 shows the percentage of the
CPU time spent on each sub-detector systems for different particle types
in a given version of the software The RICH detector took almost 50\% of
the time followed by the Ecal with about 26\%. Optical photons took the
largest fraction of the total time, followed by photons, and electrons.
Re-optimisation of the RICH modeling has already been performed reducing
the time of the RICH simulation to 20-30\% in future production
versions.

\begin{table}[bthp]
\vspace*{0.3cm}
\centering
\includegraphics[width=0.94\textwidth]{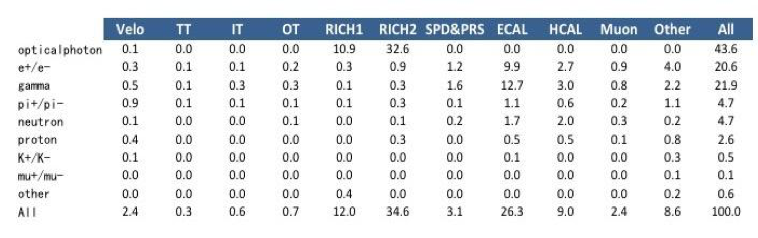}
\caption{Percentage of the CPU time spent on each sub-detector
system in LHCb for different particle types for a given version of the
software.}
\label{table:table6}
\end{table}

ALICE also consumes two orders of magnitude less CPU resources than
ATLAS. Figures 15 and 16 show charts of the percentage of time spent in
different elements of the detector simulation chain, including the
Geant4 and digitisation modules, and in the different library
components, respectively. The challenge for ALICE is to reduce the time
spent in Geant4 simulation (55\% of the total simulation time),
particularly in the geometry package and math functions, and re-optimise
the stepping algorithms. Digitisation takes a large fraction of the CPU
resources in ALICE (35\%) and is therefore also a target of ongoing R\&D
to improve the modeling of the TPC readout and the IO, and to explore
the benefits from SIMD vectorisation.

\begin{figure}[bthp]
\vspace*{0.3cm}
\centering
\includegraphics[width=0.94\textwidth]{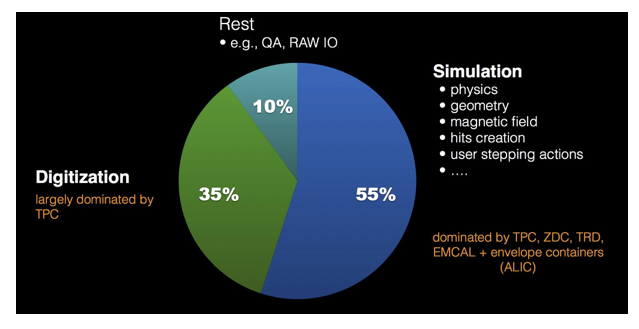}
\caption{Approximate percentage of time spent in different
components of the ALICE detector simulation chain, including the Geant
and digitisation modules (for a representative Pb-Pb Hijing event)}
\label{fig:alicecpu}
\end{figure}

\begin{figure}[bthp]
\vspace*{0.3cm}
\centering
\includegraphics[width=0.94\textwidth]{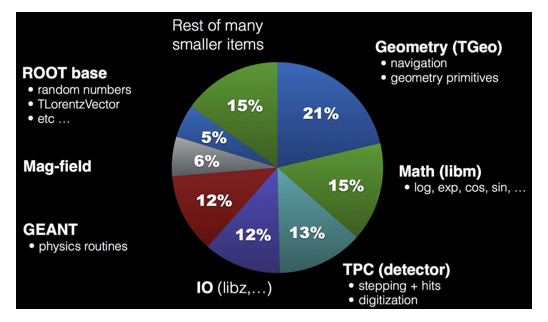}
\caption{Approximate percentage of time spent in the different
library components of the ALICE detector simulation chain for the same
setting as in Figure 16}
\label{fig:alicesimchain}
\end{figure}

\hypertarget{neutrino-experiments}{%
\subsection{Neutrino Experiments}\label{neutrino-experiments}}

Accelerator-based neutrino experiments use ``detector'' simulation tools
in two different places. Beam simulations are used to model primary
proton beam interactions with the target and the creation, integration
and decay of hadrons that produce neutrinos. For this step, most
experiments use Geant4, although some use Fluka or MARS.
Neutrino-nucleus interactions inside the detectors are taken care of by
neutrino generator tools, such as GENIE. The actual Geant4-based
detector simulation stage takes over from the generator to propagate the
final state particles through the detectors. Neutrino detectors divide
into three broad categories: segmented scintillator detectors, liquid
argon TPC's, and Cherenkov detectors.

Beam simulations generally do not require significant computing
resources. A total of 100 million events, produced at a rate of 0.01
sec/event and using much less than 1 GB per job, is generally sufficient
for a production pass. More resources are needed for assessing beam
uncertainties, simulating muons in muon monitors, and for beam design
optimisation.

For the MINERvA experiment, the whole simulation chain from neutrino
beam modeling to the production of the analysis files take about 20
sec/event, with 25-35\% of this time spent in the Geant4 module. The
required memory is typically less than 2 GB per job. The NOvA experiment
used 13 million CPU hours in 2016, 70\% of which were spent on
simulation, and less than 2 GB of memory per job.

For liquid argon TPC detectors, all Fermilab experiments use the LArSoft
software, which contains the simulation and reconstruction
packages. LArSoft includes both the Geant4 module and a ``DetSim''
module that handles the conversion from energy to ionisation charge, the
scintillator light, and the propagation of charge and photons. The
MicroBooNE experiment spends approximately 3 min/event in the Geant4
module and 1 min/event the DetSim module, with a memory footprint per
job of 4.6 GB and 2.3 GB respectively. Table 5 shows a summary of CPU
time measurements for the ProtoDUNE and DUNE Far Detector (DUNE FD)
simulation applications. Although the DUNE FD is much larger (40 kTon)
than ProtoDUNE (770 tons), ProtoDUNE will operate on the surface and
requires simulation of cosmic backgrounds. It is also not necessary to
simulate the entire 40 kTon DUNE FD in every event. For these reasons,
Geant4 simulation of the DUNE FD is substantially faster than ProtoDUNE.
Memory consumption is 2 (3) GB per job for the Geant4 module and 1.3 (1)
GB per job for the DetSim module.

In summary, computing performance is not the dominant challenge for the
simulation of scintillator-based neutrino experiments. However, the
TPC's liquid argon and segmented detector technology under consideration
for the DUNE near detector (DUNE ND) may pose significant new demands on
computing resources. The issue of liquid argon simulations requiring
large amounts of memory will be mitigated by evolving the \emph{art}
software framework ~\cite{Green:2012gv} to support multithreading.

\begin{table}[tbhp]
\centering
\begin{tabular}{| p{3.5cm} | c | c | p{2.5cm} | c | c |}
\hline
Experiment/ Simulation Stage CPU (sec/event) & Generation & Geant4 
& Geant4 with space charge & DetSim & Reco \\ \hline
ProtoDUNE (Cosmic events with Corsika ~\cite{Heck:1998vt}) & 10 & 180  & 
480 & 25 & 1000 \\ \hline
DUNE FD ($\nu$\textsubscript{e} events with GENIE) & 0.04 & 4 &   & 
44 & 160 \\ \hline
\end{tabular}
\caption{Summary of CPU time measurements for the different
modules associated with the ProtoDUNE and DUNE Far Detector (DUNE FD)
simulation applications.}
\label{table:cpudune}
\end{table}

\hypertarget{direct-dark-matter-search-experiments}{%
\subsection{Direct dark-matter search
experiments}\label{direct-dark-matter-search-experiments}}

As is the case for neutrino and muon experiments, the direct dark-matter
search experiments do not face significant computing performance
challenges. Given the typical event topology (very few tracks of limited
energy), CPU and memory consumption are not considered a particular
concern, with the exception of some specific requirements. There is,
instead, a large interest in improved and extended physics modeling.

The issues associated with physics modeling can be summarised as:

\begin{itemize}
\item
  For the liquid noble gas experiments an extension to the Geant4
  scintillation process ~\cite{1748-0221-6-10-P10002} is usually employed
\item
  For cryogenic crystal detectors the simulation of phonons and their
  interaction becomes important. Geant4 has been extended to support
  these processes ~\cite{Brandt:2014imy}
\end{itemize}

The specific CPU-related elements are:

\begin{itemize}
\item
  High-precision neutron transport may be needed in some cases. The
  associated physics models are computationally more intensive than the
  default Geant4 models.
\item
  Optical photon transportation may become a CPU bottleneck for
  experiments with a detailed and complex optical model. Advanced
  techniques leveraging GPUs are under investigation ~\cite{1742-6596-898-4-042001}.
\end{itemize}

\hypertarget{rd-towards-improving-software-performance}{%
\section{R\&D Towards Improving Software Performance
}\label{rd-towards-improving-software-performance}}

To meet the challenge of improving the performance by a large factor, an
ambitious R\&D programme is underway to investigate each component of
the simulation software for the long term. Here we will discuss the
activities that use the latest state of the art techniques to improve
computing performance. In other words, we will include R\&D efforts
other than optimisation of compiler or library settings, shower library
replacements, and optimisation of Geant4 tracking parameters or
production cuts. The use of fine grained parallelisation techniques,
novel computing architectures, and HPC systems are within the scope of
this section.

The CPU time that is spent in the simulation of LHC detectors is
typically dominated by the physics components, and in particular by the
electromagnetic part. The development of electromagnetic showers in
calorimeters plays a major role for the CPU performance of simulations.
This is also true for jets because hadronic showers, induced by the
hadron components of jets, have electromagnetic components coming from
the decays of neutral pions. The geometry component as well as particle
propagation in fields fill also important slices in the overall
performance pie-chart of simulation, as can be observed in Figure 13.
These observations are showing in which areas the R\&D effort should be
mainly focused in the short and medium term from the perspective of
detailed simulation.

Improving the CPU performance is a major milestone of simulation R\&D
and the related program of work is mainly bound to the LHC schedule for
the high luminosity phase. Adapting the simulation workflow and
algorithms to make better profit from the FLOPS potential of modern
architectures is very important. In the following we describe in detail
some of the studies to be performed in the next 3-5 years.

\hypertarget{particle-transport-and-vectorisation}{%
\subsection{Particle Transport and
Vectorisation}\label{particle-transport-and-vectorisation}}

One of the most ambitious elements of the simulation R\&D programme is a
new approach to managing particle transport, which has been introduced
by the GeantV project. The aim is to deliver a multithreaded vectorised
transport engine that has the potential to deliver large performance
benefits. Its main feature is track-level parallelisation, bundling
particles with similar properties from different events to process them
in a single thread. This approach, combined with SIMD vectorisation
coding techniques and improved data locality, is expected to yield
significant speed-ups, which are to be measured in a realistic prototype
currently under development ~\cite{1742-6596-608-1-012003}.

For the GeantV transport engine to display its best computing
performance, it is necessary to vectorise and optimise the accompanying
modules, including geometry, navigation, and the physics models. These
are developed as independent libraries so that they can also be used
together with the current Geant4 transport engine. Of course, when used
with the current Geant4 they will not expose their full performance
potential, since transport in Geant4 is currently sequential, but this
allows for a preliminary validation and comparison with the existing
implementations. The benefit of this approach is that new developments
can be delivered as soon as they are available. The new vectorised
geometry package (VecGeom), developed as part of GeantV R\&D and
successfully integrated into Geant4, is an example that demonstrated the
benefit of this approach. By the end of 2018 it is intended to have a
proof-of-concept for the new particle transport engine that includes
vectorised EM physics, vectorised magnetic field propagation and that
uses the new vectorised geometry package. This will form a sound basis
for making performance comparisons for simulating EM showers in a
realistic detector. The \emph{beta} release of the GeantV transport
engine, expected by end 2019, will contain enough functionality to build
the first real applications. This will allow performance to be measured
and give sufficient time to prepare for HL-LHC running. It should
include the use of vectorisation in most of the components, including
physics modelling for electrons, gammas and positrons, whilst still
maintaining simulation reproducibility, and I/O in a concurrent
environment and multi-event user data management.

Making use of the SIMD pipelines available today even in commodity PC's
has been already investigated at large in the GeantV project. An
important lesson resulting from this R\&D is that vectorizing the
computing intensive algorithms of the program is required but not
sufficient for enabling significant overall gains due to SIMD pipelines.
The entire simulation workflow needs to be adapted to provide the
multiple data required by vectorisation. GeantV prototyped an approach
that splits the stepping procedure for tracks into stages, aiming to
accumulate many particles before actually performing the actions
involved by the stage. As illustrated In Figure 17, the generation of
the final states from discrete physics processes can be such a
simulation stage. Particles entering it are directed to different
physics model handlers that accumulate tracks into baskets, executing
the different model algorithms in multi-particle mode. This kind of
design changes the traditional stack-based execution into a multi-stream
execution pattern. It creates the premises for increasing the code and
data locality while allowing to feed multiple data to possibly
vectorised code.

Such reshaping of the simulation has several practical implications, out
of which many are affecting directly the user code. Mixing tracks from
multiple events is required for sustaining the basketised flow. To be
noted that several current or future experiment frameworks will support
track-level parallelism as well. This will affect the data management
and I/O on the user side, requiring concurrent bookkeeping and
management of multiple events in flight. Extending the user API with
vector signatures will allow vectorised user code to exploit the
particle-level parallelism supported by the framework..

The support for an alternative approach to particle transport will also
affect the strategy for the development and long term maintenance of
physics models. An important objective of the physics developers is
therefore to review the physics algorithms and their implementation in
order to better adapt them to a multi-particle flow. This will allow a
better use of the system caches even without vectorised algorithms. For
such revisited models, GeantV is offering a testbed allowing to turn on
the basket flow and benchmark the impact of vectorisation. It is of
utmost importance that newly developed or revisited physics models are
easily ported to both types of particle transport, and this would need
to be reflected in the program of work. This will also facilitate the
validation of the new developments and benchmarking their performance in
a ``basket flow'' environment.

\begin{figure}[bthp]
\vspace*{0.3cm}
\centering
\includegraphics[width=0.94\textwidth]{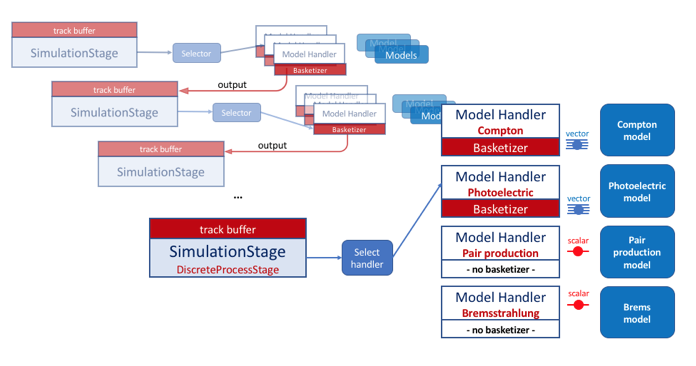}
\caption{The stepping procedure in particle transport is the
sequence of actions allowing to propagate a particle from a starting
point to the next point where the state of the particle changes
meaningfully (e.g. physics process occurrence or geometry crossing). In
GeantV, stepping is decomposed into stages, each stage selecting among a
set of models. Tracks can be buffered and sent in form of baskets to
vectorised models, which are allowed to coexist with scalar models. This
approach improves the locality and cache coherency compared to the
classical stack-based approach.}
\label{fig:parttransp}
\end{figure}

The required improvement in CPU performance cannot be achieved only from
algorithmic improvements, structural changes and multiparticle flow.
This also involves the development of fast sampling algorithms as well
as Fast Simulation approaches that bypass the simulation of full
showers. The project aims to investigate further the ways in which Full
and Fast simulation capabilities can be integrated in a single detector
simulation framework.

The initial goal is to test the transport engine with the vectorised
geometry package and demonstrate a significant speedup factor. For
simplicity, the project started by following a ``tabulated physics''
approach using values derived from Geant4, thus enabling equal footing
comparisons between the two. Tests under this approach are performed to
ensure the consistency and validity of the time performance results.
Tabulated physics is now being replaced with a physics framework using
improved and vectorised physics models, which are being validated as
they become available. Full-scale physics validation will be performed
both on available ``thin target'' data and in the context of realistic
detector applications and in collaboration with the experiments.

The vectorised components show substantial speed-ups with respect to the
corresponding code in Geant4, sometimes exceeding by one order of
magnitude. This is the case for certain solids within the VecGeom
geometry package ~\cite{1742-6596-608-1-012023}, and is planned to be extended to field
propagators and certain EM physics models. While this result is very
promising, it still does not constitute an authoritative benchmark,
although it gives confidence that the stated project goal of reaching a
significant global speedup is well within reach.

Some of the speed-up will come from optimised, vectorised physics
models, which are being developed with the objective of improving
physics accuracy and to respond to the requirements of HL-LHC and the
future intensity and energy frontier experiments. Numerous branches in
the physics code, especially in hadronics physics, pose a major
challenge to vectorisation. Since EM processes dominate showers and have
relatively fewer branches, they are the first targets of optimisation
and vectorisation. Low energy neutron and elastic hadron scattering
processes are also more amenable to vectorisation.

The estimate for the remaining work to complete the first version of the
EM physics library is of order of 1.5 FTE-years, while the work to
implement a complete first version of hadronic physics is much larger.
Current plans are to ``wrap'' an existing Geant4 hadronic physics model
to allow meaningful physics and time comparisons. This effort is
estimated at 2 FTEs.

\hypertarget{hybrid-computing-and-simulation}{%
\subsection{Hybrid Computing and
simulation}\label{hybrid-computing-and-simulation}}

Hybrid computing refers to sharing work across a mixture of computers
with different architectures. This approach seeks to exploit the
benefits of multithreading and fine grained parallelism, and needs a
software framework for writing code that executes across heterogeneous
platforms, which at the moment may include CPUs, GPUs, MICs, etc. In
detector simulation, parallelism is achieved at the level of events, as
in Geant4, or at the level of tracks, as in GeantV. Monte Carlo
techniques for detector simulations are implemented as sequential
algorithms because each \emph{step} depends on the \emph{history of
previous steps}. Although the use of these emerging technologies in
detector simulation remains a challenge R\&D efforts are to be
aggressively pursued, given the potentially large gains to be obtained.

One of the most time consuming components common to the majority of
detector simulation applications is neutron tracking, which needs to be
precise. Neutron tracking has some characteristics that make it a good
candidate to explore acceleration with the use of co-processors:

\begin{itemize}
\item
  there are many particles of the same type in the simulation;
\item
  the number of physics branches (e.g. different physics processes)
  involved in the simulation of these particles is limited;
\item
  each particle goes through a large number of steps though the detector
  material before being absorbed or leaving the detector;
\item
  when eventually an interaction occurs, the number and species of
  generated secondary particles that will further contribute to the
  signal is relatively limited.
\end{itemize}

Previous experience ~\cite{refId0} has shown that large speedup factors can be
achieved for tasks that are well-suited to GPGPU technology: simplified
geometry with only few materials, limited number of particle species and
physics processes. One example is the R\&D effort started by the Geant4
Collaboration to extend the existing code to the transport of neutrons.
One of the unique characteristics of this approach is that this system
can be integrated in existing applications, replacing only the specific
components with independent libraries. This approach is particularly
effective for HEP experiments because it allows for a non-disruptive
evolution of the simulation code that extends the thoroughly validated
current versions of the Geant4 neutron physics models (see Figure 18).

\begin{figure}[bthp]
\vspace*{0.3cm}
\centering
\includegraphics[width=0.94\textwidth]{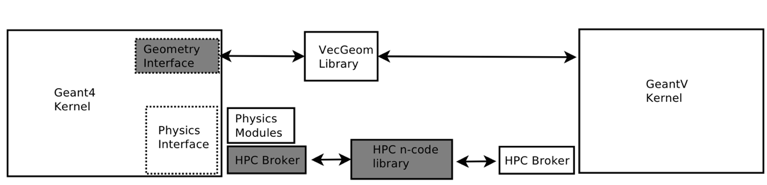}
\caption{Proposed architecture of a specialised library to
leverage GPU for neutron transport.}
\label{fig:neutrontransport}
\end{figure}

The first target of the described approach is the use of GPUs for
thermal neutron transportation. While the CPUs continue to process the
rest of the particles, the processing of all secondary thermal neutrons
would be offloaded to GPUs. Thermal neutrons propagate in the detector
and cavern for a long time before eventually interacting and generating
detector signals. Given that the latency is large, from 10 seconds to
minutes, thermal neutron depositions cannot always be associated with
the original event such that the neutron flux is the only meaningful
output information from thermal neutron transport. As granular I/O is
known to be one of the dominant obstacles against efficient use of GPUs,
a hybrid simulation with minimal I/O utilisation offers an ideal
opportunity for making efficient use of GPUs.

Along the same lines, the GeantV project achieved speedup factors of up
to 30 with respect to Geant4 when processing a basket of electrons and
photons in a GPU ~\cite{1742-6596-762-1-012019}. This is explained by the fact that EM
interactions in a detector evolve into a self-contained cascade of
electrons and photons that can exploit the benefits of vectorised code,
SIMD, and data locality. In other words, electrons and photons
traversing the same volume and getting applied the same instructions are
bundled in a single basket processed in a single thread on a GPU. The
caveat is that the baskets need to contain at least ten thousand tracks
in order for the factor of 30 to be achieved, due to the long transfer
times to and from the GPU.

\hypertarget{modularisation}{%
\subsection{Modularisation}\label{modularisation}}

Starting from the next release, a modularisation of Geant4 is being
pursued that will allow an easier integration in experimental
frameworks, with the possibility to include only the Geant4 modules that
are actually used. A further use case is the possibility to use one of
the Geant4 components in isolation, e.g., to use hadronic interaction
modeling without kernel components from a fast simulation framework. As
a first step a preliminary review of libraries granularity is being
pursued, which will be followed by a review of intra-library
dependencies with the final goal of reducing their dependencies. By
2019, it is intended to redesign some Geant4 kernel components to
improve the efficiency of the simulation on HPC systems, starting from
improved handling of Geant4 \emph{database}s on large core-count
systems. A review will be made of the multithreading design to be closer
to the task-based frameworks, such as Intel's TBB.

\hypertarget{simulation-on-high-performance-computing-systems}{%
\subsection{Simulation on High Performance Computing
Systems}\label{simulation-on-high-performance-computing-systems}}

HPC resources at supercomputer centers can provide a substantial
fraction of the needed computing power to address HL-LHC needs. The
Geant4 Collaboration has started an R\&D program of code optimisation
for a more efficient utilisation of HPC resources. One of the main
features of the next generation of supercomputers is the use of
massively parallel architectures that require the combination of novel
techniques to achieve scalability. The release of multi-threaded Geant4
in 2013 has shown that it is possible to obtain almost perfect linear
scaling, whilst keeping memory usage under control. Subsequent research
has focused on extending the parallelism to multi-node with HPC-friendly
techniques (MPI), using a mixed MT/MPI approach. A recent result has
shown that scaling for hundreds of thousands of concurrent threads can
be achieved for a semi-realistic HEP-Geant4 application. A test
performed on the MIRA supercomputer at ANL successfully ran a modified
version of Geant4 spawning more than 3 million threads. The
test concluded that the main bottleneck to achieve scaling linearity is
concurrent access to I/O resources (see Figure 19). Extensions to Geant4
capabilities to improve the scalability of the code are underway. These
include:

\begin{itemize}
\item
  parallel merging of scorers/histograms and ntuples via MPI;
\item
  redesign of RNG Seeding in large parallel applications;
\item
  investigation of use of HPC-friendly data-storage for input data-base
  libraries.
\end{itemize}

A custom version of Geant4 developed for tests on MIRA showed
scalability for up to 3 million threads. This version is the basis for
further development to be included in future Geant4 public releases. A
spinoff result from the MIRA tests has been the validation of Geant4
physics to 1/10\textsuperscript{5}, a level of precision never achieved
before.

ATLAS is actively pursuing HPC resource utilisation to supplement grid
resources for simulation production. Multi-process jobs distributed with
YODA have been running successfully in production on supercomputers at
NERSC and machine utilisation is being studied ~\cite{1742-6596-664-9-092025}. ATLAS is
currently migrating its simulation application to multi-threading to
ensure good scaling performance on HPC architectures with minimal memory
footprint. Current measurements of multi-threaded event throughput and
memory consumption scaling on Intel Xeon Phi Knights Landing chips are
very promising, demonstrating ability to scale well up to the maximum
number of threads on a device with a factor five decrease in the
per-worker memory consumption relative to multi-process jobs ~\cite{Farrell:2242857}.
Overall throughput performance falls short of traditional Xeon
architectures, however, so studies are ongoing to identify and remove
the performance bottlenecks.

CMS has also been studying how best to exploit HPC resources by running
combined multi-step jobs that cover Generation, Simulation, Digitisation
and Reconstruction in a single workflow. The results when scaling to
high-pileup scenarios have shown that limitations are primarily due to
the reconstruction step, rather than the simulation step. Studies have
been made of various ways of filling a whole KNL many-core node with
Generation and Simulation jobs of varying sizes and the highest total
node event throughput was achieved with two jobs using 128 threads each.
The simulation application and its underlying simulation engine, Geant4,
is not limiting scaling behaviour.

\begin{figure}[bthp]
\vspace*{0.3cm}
\centering
\includegraphics[width=0.94\textwidth]{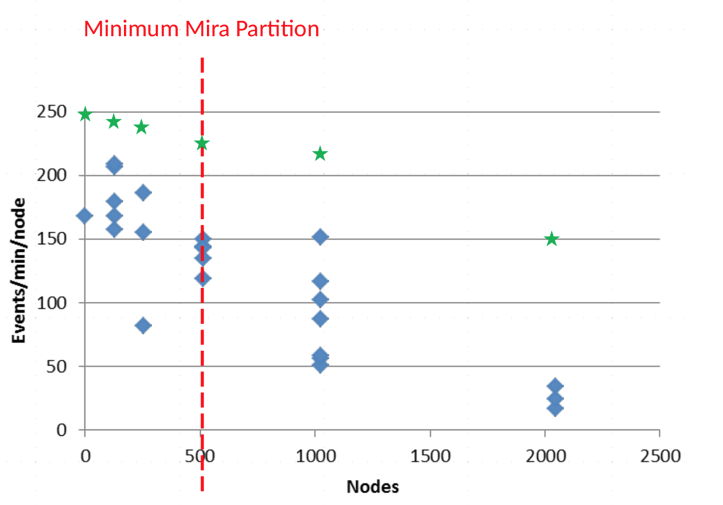}
\caption{Top: Initial testing of Geant4 on MIRA (blue diamonds) and VESTA
(green stars). Each node has 80 threads. The loss of scalability is due
to I/O and intra-node communication limitations (MIRA nodes do not have
disks and rely on MPI for every I/O operation). VESTA, a smaller
test-bed, shows better results because the number of I/O nodes is
larger. }
\label{fig:mira}
\end{figure}

\hypertarget{fast-simulation}{%
\section{Fast Simulation}\label{fast-simulation}}

As pointed out earlier, simulation consumes the majority of the CPU time
used by many experiments. Optimisation studies on detailed simulation
codes are ongoing and can provide helpful performance improvements, but
in order to satisfy the needs of an entire experiment, other options
must be explored. The increase in size of the MC samples needed to match
those of the data collected by upgraded or future experiments, as well
as the increase in the complexity of the events, will put additional
strains on the CPU needs. While the availability of an accurate detailed
simulation will remain of paramount importance, often the full
high-fidelity description of the detector implemented by the experiments
is not necessary. There are many cases, including detector upgrade
studies, tests for evaluation of systematic uncertainties in physics
measurements, scans of large signal phase spaces in the context of new
theories, and prototyping of new analysis ideas, where the demand on
simulation physics accuracy may be relaxed. Even within a given physics
analysis, it may be good enough to evaluate some quantities with high
accuracy in a small MC sample and other quantities with lower accuracy
in larger samples. A combination of different fast simulation approaches
targeting parts or the whole simulation application are the optimal
solution for saving computing resources in these cases.

In general, simulation time scales with the energy of the particles
being simulated, and with the minimum energy down to which they are
simulated. This means that by far the most time is taken by showers in
calorimeters, where large numbers of low-energy electromagnetic
particles are produced and propagated by the simulation. Because the
simulation time scales with energy, rather than transverse energy, the
time consumed by LHC simulations is dominated by the forward regions,
which are the target of the highest energy particles in most collider
events. Fast simulations options for calorimeter showers have the
side-effect of reducing significantly the number of low energy neutrons
in the simulated event, which contribute significantly to the simulation
time due to their relatively long lifetimes. Particle propagation in
tracker detectors is typically the next most time-consuming step in a
simulation. This step consists mostly of charged particle propagation,
and the geometry detail is the most important aspect to be addressed by
the a fast simulation. Heavy Flavour physics experiments utilise Ring
Imaging Cherenkov (RICH) or Detection of Internally Reflected Cherenkov
light (DIRC) detector technology. The transport of Cherenkov photons
through optical processes needs to be simulated in detail in these
cases, and requires more CPU time than regular trackers. Fast simulation
of these processes is in general tied to the reconstruction technique
used in a given detector.

There are two main types of fast simulations. One is the replacement of
a particular physics process or set of processes with a module that
contains faster code. The other approach is the simulation of an entire
particle trajectory or shower, or even a more complex object like a
hadronic jet, with a faster software module. These faster modules often
consist of parameterised models of the physics processes or particle
trajectories. While the technique could be shared, the exact
implementation would be different for each experiment. Much of this
section focuses on the second technique, while the first is discussed
further in Section 10.3 on machine learning.

It is worth noting that the fast simulation techniques and software have
relevance outside the HEP domain. For example, radiation transport
techniques are extremely important for a wide range of applications:

\begin{itemize}
\item
  radiation protection in all industrial fields using ionizing
  radiations (sterilisation, imaging etc.);
\item
  radiation protection for space flights;
\item
  design and optimisation of imaging medical instruments;
\item
  treatment planning for radiation therapy;
\item
  radiation safety for nuclear power and fuel processing plants;
\end{itemize}

All these applications would greatly benefit from an increase in the
simulation speed with important benefits to their respective fields.

A way to achieve an additional speedup is to customise the simulation to
the needs of a specific physics analysis. This would allow to target a
given sub-detector or part of the event.

\hypertarget{fast-simulation-techniques-and-examples}{%
\subsection{Fast Simulation techniques and
examples}\label{fast-simulation-techniques-and-examples}}

There are many techniques to make simulation faster. Some of them are
applied to Geant4-based Full Simulation applications, such as for
example Russian Roulette and shower libraries, to speedup simulation in
a given sub-detector. These applications are still
labeled Full Simulation and Geant4-based. Some experiments have
developed alternative fast simulation frameworks, based on fast track
propagation through a simplified geometry and parameterised showers,
which are less accurate and are typically used in detector upgrade
studies and to produce ``signal'' samples that require scans of a large
region of theory parameter space. Generally, the faster the simulation,
the more imprecise the physics is. Examples of fast simulation
techniques, used to speedup Geant4-based applications or in fast
simulation frameworks, are:

\begin{itemize}
\item
  \textbf{Russian Roulette:} Some fraction of the simulated particles
  are discarded and a weight is applied to the others to compensate for
  this effect. It is commonly used in the simulation of thermal
  neutrons, which have long lifetimes and consume a significant amount
  of CPU time while propagating through matter before being absorbed.
  Russian Roulette is used in CMS's Full Simulation application and
  yields 25\% (30\%) CPU savings when processing minimum bias (ttbar).
\end{itemize}

\begin{itemize}
\item
  \textbf{Shower libraries:} A particle of a given type and in a given
  energy range and geometry region is replaced with the simulation
  information of a similar particle stored in a library of showers
  pre-simulated with a detector simulation tool such as Geant4. Shower
  libraries are typically used to replace Full Simulation in
  sub-detectors with high particle occupancy, such as forward detectors
  in colliders. The are efficient to model high-energy showers where a
  large number of low-energy secondary particles are produced. If
  applied to the entire detector, they achieve speedup factors of 2-5.
  Both in ATLAS and CMS, shower libraries are used by default in the
  forward hadronic calorimeters, yielding a 25\%-40\% time performance
  gain in the case of CMS and a 30-50\% performance gain in the case of
  ATLAS. ATLAS tests indicated previously that extending the use of
  these libraries to the remainder of the calorimetry could bring gains
  of an additional 50\%. LHCb is currently exploring its use.
\item
  \textbf{Parameterisation of calorimeter showers}: The simplest
  implementation of a parameterised particle shower involves the
  parameterisation of the shower shapes of high- and low-energy EM
  particles. In sampling detectors with complex geometries, these
  parameterisations are difficult to tune. Additionally, with a
  relatively thick inner detector, and in the case of ATLAS a cryostat
  and solenoid in front of the active calorimeter region, there may not
  be many high-energy electromagnetic particles available to the fast
  simulation. This simplest approach, based on Ref. ~\cite{Sjostrand:2014zea}, was
  abandoned by ATLAS about ten years ago. Current approaches used by
  both ATLAS and CMS involve more complex modeling of the showers than
  can be achieved by a simple parameterisation. Although not part of the
  current default simulation, CMS implemented shower parameterisations
  using GFLASH to model the shower after the first interaction
  of each primary particle with matter, by tuning a set of parameters to
  fit the geometric distribution of the associated energy depositions
  obtained from a Geant4 simulation. The ATLAS approach uses histograms
  with energy deposit and shower shape information to directly treat
  particles before they enter the cryostat. All these approaches have
  the advantage that they can ignore, during energy deposition, the
  detailed geometry of the calorimeter, and deal only with the readout
  geometry. All have the option of building objects similar to the
  energy depositions from Geant4 or attempting to directly build
  summary-like objects that are used in digitisation. Typically,
  parametric simulation achieves speedup factors of 100-1000 in the part
  of the detector treated by the parameterisation.
\item
  \textbf{Low energy background parameterisation}: Muon detectors are
  located behind the calorimeter system preventing most high energy
  particles of a different nature to reach them. The main contribution
  to the occupancy in these detectors, apart from muons, is due to
  leakage from calorimetric showers and thermal neutrons. In order to
  obtain fully realistic occupancies in the muon system, photons,
  electrons and neutrons need to be transported to very low energy, and
  the surrounding infrastructure and accelerator elements must be
  modeled. In the LHCb experiment, the addition of this detail to the
  full Geant4 simulation would result in a CPU time increase of the
  order of a factor of 10. Instead, this low energy background is
  simulated via a parameterisation of the difference between a Full
  Simulation with and without the low energy processes. A similar method
  is being validated by the ATLAS collaboration.
\item
  \textbf{Fast tracker simulation}: Both ATLAS and CMS have custom fast
  track simulation algorithms based on simplified geometries. The time
  consumed by the Full Simulation of a tracking detector is primarily
  spent on the propagation of charged particles in a magnetic field,
  which is directly related to the number of times the field must be
  sampled. By simplifying the geometry, for example by considering
  infinitely thin layers, the number of samples is significantly
  reduced, resulting in a faster overall simulation. The other way to
  speed up the tracker simulation is by introducing fast, parameterised
  models for material interactions, particularly for the more common
  electromagnetic interactions. The fast track simulation code used by
  ATLAS has been made public as a part of the common tracking software
  project ACTS ~\cite{ACTS}, adopted by the FCC. These fast track
  simulations offer speed improvements of 10-1000x over a detailed
  simulation, though for a Fast Simulation including both fast tracking
  and fast calorimeter simulation, tracking is typically the dominant
  resource consumer.
\item
  \textbf{Fast digitisation and reconstruction}: Especially for higher
  pileup at the LHC, once fast calorimeter simulation is used, and
  certainly when fast tracker simulation is included, digitisation and
  reconstruction become the dominant consumers of CPU time in the
  simulation chain. As a result, both CMS and ATLAS have implemented
  fast modules for specific pieces of the digitisation and
  reconstruction code that are most time-consuming, primarily in fast
  inner detector digitisation and fast tracking. Fast tracker hit
  reconstruction uses template hit smearing functions derived from full
  reconstruction, whereas fast tracking often uses generator (truth)
  particles as seeds to avoid the combinatorial issues of standard track
  reconstruction software. The speed improvements from these algorithms
  varies with pileup but are generally more than one order of magnitude
  ~\cite{1742-6596-523-1-012035}.
\item
  \textbf{Fully parameterised simulation:} Many programs are available
  in the public domain for the implementation of very fast parameterised
  simulation, the most popular being DELPHES ~\cite{deFavereau:2013fsa}
  and PGS ~\cite{PGS4}. These
  tools are extremely fast, running at hundreds of events per second,
  but they lack the accuracy of the alternatives discussed before. These
  programs are heavily used by phenomenologists for re-interpretation of
  LHC results, and by ATLAS, CMS, and LHCb for upgrade detector
  performance studies.
\item
  \textbf{Ultra fast self-tuning non-parametric simulations:} A
  re-emerging alternative trend in simulation is to develop ultra fast,
  self-tuning simulators based on lookup tables that directly map
  generator events into simulated events. The lookup tables in this case
  are formed directly from fully simulated events. A past example was
  Turbosim ~\cite{Knuteson:2004nj} developed at the Tevatron for the D0 and CDF
  experiments. A prototype for a redesign called Falcon ~\cite{Brooijmans:2016vro} was
  recently released, with a more efficient algorithm for the
  multi-dimensional mapping of particle properties.
\end{itemize}

A natural evolution of having separate Full and Fast Simulation
frameworks in experiments is the concept of ``hybrid'' simulation (see
Section 9.2), where a single framework allows to treat some aspects of the
event with a fast or very fast simulation technique and others with a
detailed simulation approach. Such configurations might be useful for
handling pile-up and also for B-physics, where only a few particles are
of great interest to the analysis while it is helpful to have the others
simulated with some minimal accuracy. The difficulty with the hybrid
approach is that the optimal configuration is analysis dependent and
requires the derivation of customised calibration and efficiency
factors, activities which are person-power intensive.

LHCb has developed simulation code for reuse of the underlying event in
the case where only a few particles are of interest and the contribution
of the rest only degrades the `signal' measurement. The technique
consists of fully simulating the underlying event once, while the
`signal' is simulated N times independently and recombined with the one
underlying event. It allows a reduction in CPU time consumption
proportional to the number of times the underlying event is reused.

\hypertarget{physics-performance-of-fast-simulations}{%
\subsection{Physics Performance of Fast
Simulations}\label{physics-performance-of-fast-simulations}}

The fastest of the fast simulations are typically useful for situations
in which only approximate answers are needed, or where large
uncertainties are not a serious concern. Such situations are
surprisingly common. For example, studies of BSM limits on new particles
in LHC upgrade scenarios, where cross-sections fall rapidly with mass,
can afford a 100\% uncertainty on the analysis efficiency as it results
in a relatively small uncertainty on the excluded mass. The slower
simulation techniques generally provide higher fidelity, but they are
more resource intensive.

The performance of Fast Simulation in ATLAS, CMS, and LHCb is evaluated
with respect to the Geant4-based Full Simulation. For the fastest
version of ATLAS FastSim, and for CMS FastSim, the agreement is on
average quite good for kinematic distributions such as the momentum and
pseudorapidity of high level event objects, such as light jets, b jets,
missing transverse energy, muons, electrons, photons and taus. Larger
discrepancies are observed when comparing variables that depend on the
structure or shape properties, such as jet particle composition or
variables used for electron and photon identification. Fast simulations
have difficulty reproducing the detailed structure of hadronic shower
shapes, which create issues in modelling boosted objects and jet
substructure, which are very relevant signatures at higher energies.
Another common issue is in the treatment of exotic particles. Particles
with high charge (e.g. Q-balls), strange propagation (e.g. Quirks),
unusual nuclear interaction patterns (e.g. R-hadrons), and almost any
particles with late decays that result in displaced vertices or
disappearing tracks, for example, are difficult to treat with fast
simulations because of the assumptions built into the parameterisations.
For example, most fast simulations assume that particles are pointing,
while particles from late-decaying heavy particles may not be.

Fast simulations do have the advantage that they are generally more
tuneable than a detailed simulation toolkit like Geant4. The knobs for
tuning are generally better exposed to users (particularly for cases
like PGS and DELPHES), and they have a more clear connection with
physical observables. Thus, it can be easier to tune a fast simulation
to match the data than it is to tune a detailed simulation toolkit.

Figure 20 shows the speed of CMS Fast Simulation factored into different
subprocesses as a function of the number of pileup interactions. The
sub-processes are defined as follows:

\begin{itemize}
\item
  \emph{Trajectory sim.} refers to fast simulation of the particle
  trajectory through the detector, and formation of tracker and muon
  simulated hits.
\item
  \emph{Calorimeter sim.} refers to fast calorimeter simulation.
\item
  \emph{Tracker hits} refers to a FastSim specific simplified hit
  reconstruction in the tracker.
\item
  \emph{Pileup mixing} refers to the process of mixing hits from a
  physics process of interest with those from a minimum bias sample
  representing pileup.
\item
  \emph{Tracking} is the MC truth based, fast track finding algorithm.
\item
  \emph{Digitisation} is the simulation of readout, which is the same in
  fast and Full Simulation in CMS.
\item
  \emph{Calorimeter reco.} is the reconstruction of calorimeter hits,
  which is standard for all simulation and data.
\item
  Muon reco (not shown in the plot): is the reconstruction of muon
  detector hits, and takes on average \textasciitilde{}0.01s/event.
\item
  Finally, \emph{Object reco.} is the standard CMS reconstruction of all
  objects including vertices, jets, electrons, photons, muons and taus
  used in data and simulation.
\end{itemize}

\begin{figure}[bthp]
\vspace*{0.3cm}
\centering
\includegraphics[width=0.94\textwidth]{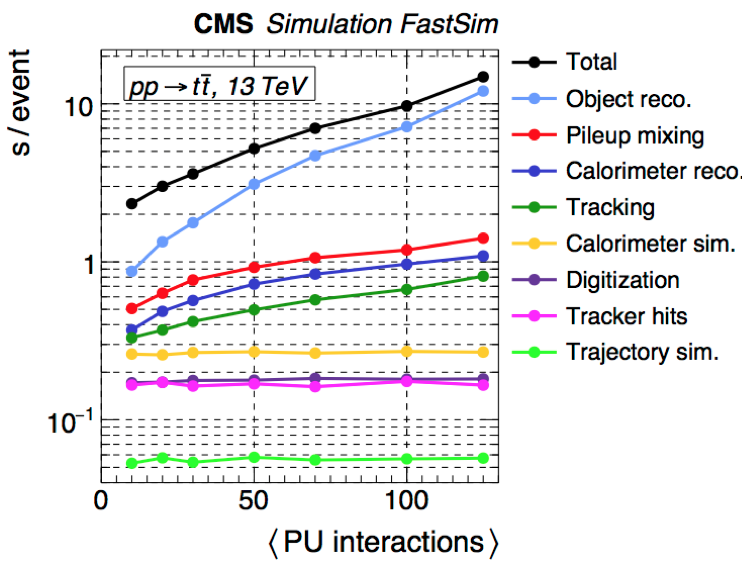}
\caption{Speed of CMS Fast Simulation factorised into different
subprocesses, as explained in Section 10.2, versus number of pileup
interactions. }
\label{fig:speedfastsim}
\end{figure}

\hypertarget{machine-learning-in-fast-simulations}{%
\subsection{Machine Learning in Fast
Simulations}\label{machine-learning-in-fast-simulations}}

There has recently been a great deal of interest in the use of machine
learning in Fast Simulation, most of which has focused on the use of
multi-objective regression ~\cite{2007physics...3039H} and generative adversarial networks
(GANs) ~\cite{2014arXiv1406.2661G},~\cite{deOliveira:2017pjk},
~\cite{Paganini:2017hrr}. Since use of GANs allows for non-parametric
learning in cases such as calorimetric shower fluctuations, it is a
promising avenue for generating non-Gaussian and highly correlated
physical effects. This is an obvious area for future expansion and
development, as it is currently in its infancy. Just as with Fast
Simulation in general, one can imagine two types of machine learning
application to simulation. One is the replacement of a particular
physics process, or set of processes, with a machine learning module.
Such an approach has the clear advantage that it can be shared among the
experiments as a part of a general simulation toolkit like Geant4. The
approach has the disadvantage that, while it can be used to directly
replace a specific module of the toolkit by training against the
standard model, it otherwise might be rather difficult or cumbersome to
produce training data that can be used for the development of a clearly
distinct new model of physics. The other approach is the replacement of
a fast simulation technique with a machine learning module - that is, a
machine learning module that performs the entire simulation of a
particle, or even a more complex object like a hadronic jet. Machine
learning approaches also offer the opportunity for end-to-end simulation
where the output can represent composite objects resulting from multiple
simulation steps. While the technique, and perhaps even some details of
the network and training methods, could be shared, the exact
implementation would most likely have to differ among the experiments.
Discussions and sharing involved in this latter scenario is more
appropriate in an inter-experiment ``forum'' like the recently developed
Inter-experiment Machine Learning Working Group at CERN ~\cite{IML}. The
sharing of code and other technical details, in addition to ideas and
results, is essential in order to expand the benefits to the entire
community.

Most current work is focused on using multi-objective regression and
generative networks on simple events, such as single particle
calorimeter showers. Even though these are the most basic events to
simulate, more complex observables can be built as straightforward
combinations of single particle showers. In all cases, one must choose
an output, whether a general type like an energy deposit that might be
common to many experimental simulation applications, or a specific type
like the detector response that would presumably be applicable to only
one experiment. The technique and methodology, however, can be shared
among experiments even when their output formats differ.

The applications of machine learning are numerous, and the coming period
will involve a detailed exploration of many different approaches to
improving or speeding up the simulation using machine learning tools.
Within ATLAS, for example, neural networks are being explored as a
replacement for the histograms used in fast calorimeter simulation
mentioned earlier. This type of machine learning tool could also be
applicable to the storage of cross-sections within Geant4, for example.

While these techniques are quite promising, these are early days, and
detailed speculation about the adoption of machine learning across the
field is difficult. The fact that machine learning tools like
SciKitLearn ~\cite{DBLP:journals/corr/abs-1201-0490} and 
TensorFlow ~\cite{DBLP:journals/corr/AbadiABBCCCDDDG16} are being developed by
professionals and in an open-source community means that if the
simulation community takes full advantage of these publicly available
tools then the optimisations and developments of the broader community
will be immediately useful. Already, the parallelism of standard machine
learning tools used to implement GANs have made it possible to foresee
significant improvements from parallelism and vectorisation, even with
current simulation frameworks based on Geant4.

The first neural network-based fast simulations are now publicly
available, along with code, documentation, and training data. 
These first results are very promising and establish a
baseline for continuing to improve the performance in preparation for
integration into real experimental settings. Dialog between the
experiments has been, and will continue to be, important to push these
techniques forward.

There are two related approaches for replacing or augmenting fast
simulations with generative models. One approach is to construct
tailored architectures that are specifically designed for all or part of
a particular detector geometry and detector response. Another approach
is to develop generic tools that can be adapted for universal or common
aspects of simulation in many experiments. It is likely that a synergy
of both approaches will be required to extract the most performance from
state-of-the-art methods. The GeantV collaboration already has an active
program to develop generic tools for this second approach. Within the
GeantV project, it is planned to develop a general interface where a
user can specify a physics process and detector type and get back an
optimal trained machine-learning model that will simulate the detector
response. Such a tool, once developed and validated, will be usable with
any transport Monte Carlo, including Geant4 with its well-developed fast
simulation interface. The GeantV R\&D program also includes tests of the
generative neural networks in the context of Geant4. The tool will
incorporate various trained machine learning models, predictive
clustering trees and generative neural networks. Early studies have
shown that choosing heuristically a network layout and then training it
may not lead to the expected results. The fundamental parameters of the
network need to be adapted to the problem under study. This opens the
possibility of introducing some of the parameters defining the network
structure among the quantities to optimise in the training process.
Although this moves the whole optimisation process one step further by
introducing a meta-optimisation that can be accomplished in the same
step or in alternate steps to the more traditional network training, it
offers the possibility to go beyond the one-size-fits-all in Machine
Learning. This will remove part of the arbitrariness in choosing a given
network structure, and will provide the opportunity to control the
trade-offs between accuracy of the results and training and computation
time. In ML parlance, this is called an embedded algorithm, where the
information from the feature performance is fed back to the algorithm
building stage together with the meta-optimisation, usually called
\emph{hyper-parameter} \emph{tuning}, leading to problem-specific
optimised classifiers. If successful, this could provide an opportunity
for shared not only techniques but also work among the experiments, as
the meta-optimisation could account for some of the differences between
detectors and targets.

While the focus in this section and in the literature so far has been on
using machine learning to do fast simulation (e.g. learn Geant4), it may
be possible to train a generator directly on data to replace or augment
Full Simulation. This would require high fidelity test beam data for a
variety of particles and energies. While this is beyond the scope of
this CWP, we suggest that this be pursued in parallel to improving the
speed and accuracy of fast simulators.

\hypertarget{sharing-of-fast-simulation-and-interfaces}{%
\subsection{Sharing of Fast Simulation and
Interfaces}\label{sharing-of-fast-simulation-and-interfaces}}

At the moment, simulation toolkits like Geant4 provide interfaces for
the experiments to ``hook in'' their fast simulation code. These have
proven sufficient for very simple fast simulations. Because many aspects
of Fast Simulation depend on experiment specific issues (not just in
terms of detector technologies, but in terms of actual C++ types used,
as in many cases intermediate classes are bypassed to save time), it can
be difficult to find commonalities. However, the techniques, approaches,
and methods, as well as some of the issues encountered during
implementation or understanding of the tuning of Fast Simulation, can be
common among the experiments. A forum in which such issues could be
discussed would be very useful, particularly as machine learning
techniques become more widely adopted.

In GeantV, Fast Simulation is being integrated as a user process,
offering the same flexibility as in Geant4 to combine fast with Full
Simulation. The Fast Simulation processes will get as input the full
track state information, allowing to limit the step length and therefore
select particles for performing the user-defined Fast Simulation based
on arbitrary selection criteria, such as particle type, energy,
detector, etc. As in the case of Geant4, it will be possible to have
physics lists per region, which in case of Fast Simulation gives extra
flexibility for implementing some use cases. The main difference however
will be the possibility to put Fast Simulation in the multi-particle
vectorised flow. As any physics model, Fast Simulation will provide not
only a scalar interface (allowing the user code working with Geant4 to
work also with GeantV), but also multi-particle interfaces. The Fast
Simulation user models that are vectorised will be able to benefit from
data and code locality, as well as instruction level parallelism as any
other component of GeantV, while the scalar models will be still usable.

\hypertarget{pseudorandom-number-generators}{%
\section{\texorpdfstring{\textbf{Pseudorandom number
generators}}{Pseudorandom number generators}}\label{pseudorandom-number-generators}}

The use of sequences of pseudo-random numbers is critical for all Monte
Carlo simulations. In simulating particle and radiation transport the
use of pseudorandom number generators is a vital component. The large
number of random numbers required in the simulation of LHC and other
High Energy and Nuclear Physics (HENP) detector events, as well as
simulation in other fields, and the need to obtain statistically correct
estimate while simulating with a large or very large degree of
parallelism present challenges in the selection and use of pseudorandom
number generators (PRNGs).

Reproducibility is a key requirement of HENP simulation i.e. when
repeating the simulation of an event with the same input particles, with
the same state of a PRNG (in the same geometry and with the same physics
models), the simulation must provide exactly the same results, i.e. the
same values for all observed quantities. Moreover it is required because
it allow problems to be located and debugged. It enables the
verification of new experiment software versions when based on the same
simulation release (e.g. Geant4 release 10.3 patch 1) and even of new
hardware, given the same binaries.

The results must be statistically equivalent for all PRNG, and obtaining
different results with any PRNG is a clear signal that at least one of
the PRNGs is not adequate for the simulation. Correct results require
that the sequences do not contain correlations.

Sequential runs of detector simulation in one process, e.g. using
Geant4, only require a single initialisation of a PRNG at the start of a
simulation run. The state of the PRNG at the start of an event can be
recorded, and stored with the event, to enable its reproduction.

In event-parallel multi-threaded simulation, such as multi-threaded
Geant4, reproducibility means that each event must be provided with a
deterministic seed (or state), typically created at the start of the
simulation. This method is a simple extension of event seeding used in
experiment grid productions.

Nevertheless only some PRNGs offer guarantees for the correlation of
such sequences. Thus tests to ensure that all simulation results are
unaffected by changing the type of PRNG engine is strongly advised. This
use is similar to the research on parallel PRNG for use in other fields
~\cite{PhysRevE.75.066701}. A recent survey ~\cite{lecuyer:hal-01399537} covers 
developments for parallel PRNGs, including ones for GPUs.

When each track of an event can be simulated in different order, as in
the case of the GeantV simulation design, the state of the PRNG must be
attached to a track. This requires memory proportional to the size of
the state of the PRNG and to the number of tracks in flight. This
excludes the use of PRNGs with large state. A number of potentially
suitable PRNG classes have been identified, including counter-based
PRNGs, such as Random123 ~\cite{Salmon:2011:PRN:2063384.2063405}, 
Combined Multiple Recursive Random number Generators (CMRGs) 
~\cite{doi:10.1287/opre.47.1.159} and the newest extension of the
MIXMAX PRNGs ~\cite{Savvidy:2015jva} with small matrix size.

These new families of PRNGs are largely untested in radiation transport
simulation, yet are at the forefront of PRNG 'technology' for use in
parallel applications. Thus a key aim should be to obtain
implementations compatible with current interfaces (CLHEP and ROOT) so
that they can be tested in large scale simulations, before their use
become necessary in the transition to large-scale or fine-grained
parallelism.

One of the methods for seeding parallel threads is to create
sub-sequences from a single PRNG with a very long period. The mechanism
for choosing a subsequence, whether it depends only on the sequencing of
tracks or on additional transport parameters, and its computational cost
are areas of active investigation. The resulting algorithms to generate
a new state for each daughter track, must be validated in trial
simulations.

In addition, the new C++11 \textless{}random\textgreater{} interface and
its implementations must be considered. These have not yet seen
widespread use in simulation. Trials within the Root project raised an
issue with the performance of distributions that output floating point
numbers (single and double precision) for several popular generators. If
these findings are confirmed, we must seek to create faster, lower
precision, distributions for floating point numbers for particular
generators. In current practice a single function call is used to obtain
many variates (output numbers) in performance critical code.
Investigations will be made as to whether it is possible to create
similar functionality by designing and implementing a new type of
distribution object, or obtain similar performance benefits in a
different way.

The future programme of work in the field of PRNGs includes the
following:

\begin{itemize}
\item
  Ensure that performant implementations of additional state-of-the-art
  PRNGs are made available, in particular PRNGs which have been shown to
  have none or minimal correlations between separate sequences or
  subsequences.
\item
  Promote the replacement of obsolete PRNGs currently in use in HEP
  experiments, including HepJamesRandom, RanECU, whose periods are small
  by the standards of today's computing power, and the number of output
  numbers required in typical event simulations.
\item
  Create vectorised implementations of PRNGs for use in sequential,
  event-parallel and track-parallel (fine grained) simulations.
\item
  Develop a single library containing sequential and vectorised
  implementations of the set of state-of-the-art PRNG, to replace the
  existing Root and CLHEP implementations within 3 years. This includes
  counter-based methods, the MIXMAX family of generators, improved
  implementations of RANLUX, implementations of CMRGs and other
  categories of sequential and parallel PRNGs. Promote a transition to
  the use of this library to replace existing implementations in ROOT,
  Geant4 and GeantV.
\item
  Collaborate with the authors of the state of the art PRNG testing
  suites, and in particular TestU01~\cite{Ecuyer:2007:TCL:1268776.1268777}, 
  encouraging and seeking to contribute to the extension to increase 
  the statistics of testing, extend testing to 64-bit variates, 
  and expand the testing of correlations between sub-sequences.
\item
  Follow and contribute to the evolution of the field of PRNGs for
  parallel and highly parallel applications, collaborating with
  researchers in the development of PRNG, seeking to obtain generators
  which address better our challenging requirements.
\end{itemize}

\begin{itemize}
\item
  Investigate whether poor PRNGs, or PRNGs with few, known deficiencies
  can create erroneous results in the particle-transport simulation of
  simple setups.
\end{itemize}

\hypertarget{summary-and-outlook}{%
\section{Summary and outlook}\label{summary-and-outlook}}

In this note we have attempted to describe the main challenges and
opportunities for Full and Fast Simulation applications of HEP
experiments in view of planned future experimental programs in the next
decade; these include the HL-LHC, neutrino and muon experiments, and
studies towards future colliders such as FCC and CLIC. The combination
of additional physics models and higher accuracy in the simulation with
the need of larger Monte Carlo samples, which scale with the number of
real events recorded in future experiments, place an additional burden
on the computing resources that will be needed to generate them.

A multi-prong approach is being followed as a strategy for responding to
these challenges. Firstly Geant4 will continue to deliver new or refined
functionalities both in physics coverage and accuracy, whilst
introducing software performance improvements whenever possible. At the
same time, an R\&D programme is investigating new approaches that aim to
benefit from modern computing architectures. For example, 
the main feature of the GeantV transport engine is
track-level parallelisation, bundling particles with similar properties
from different events to process them in a single thread. This approach
combined with SIMD vectorisation coding techniques and use of data
locality is expected to yield large speed-ups, which are being measured
in a realistic prototype under development. In addition, the work on
Fast Simulation is accelerating with a view to producing a flexible
framework that permits Full and Fast simulation to be combined for
different particles in the same event. The overriding requirement is to
ensure the support of experiments through continuous maintenance 
and improvement of the Geant4 simulation toolkit with minimal
API changes visible to these experiments at least until production
versions of potentially alternative engines, such as those resulting
from ongoing R\&D work, become available, integrated, and validated by
experiments. The agreed ongoing strategy to meet this goal is to ensure
that new developments resulting from the GeantV R\&D programme,
including the novel transport engine, are
deployed in a timely way in Geant4.

Improving the speed of simulation codes by an order of magnitude
represents a huge challenge. However the gains that can be made by
speeding up critical elements of the common Geant4 toolkit can be
leveraged for all applications that use it and therefore will have a big
impact on computing resource requirements. It is important to note that
the current effort level available in the community is barely sufficient
to keep up with the maintenance and improvements required by current and
future experiments. It is therefore of critical importance that the
whole community of scientists working in the simulation domain continue
to work together in as efficient way as possible in order to deliver the
required improvements. Very specific expertise is required across all
simulation domains, such as the physics modeling, tracking through
complex geometries and magnetic fields, and building realistic
applications that accurately simulate highly complex detectors.
Continuous support is needed to recruit, train, and retain the
person-power with the unique set of skills needed to guarantee the
development, maintenance, and support of simulation codes over the
short, medium, and long timeframes foreseen in the HEP experimentals
programmes.

\sloppy
\raggedright
\clearpage
\printbibliography[title={References},heading=bibintoc]

\end{document}